# Orientation of Entrance and Burial Chamber in the Pyramids of the Egyptian Fourth and Fifth Dynasties

Amelia Carolina Sparavigna

Department of Applied Science and Technology, Polytechnic University of Turin, Italy

In a recent work, arxiv:2412.20407, we have mentioned two Egyptian kings, Shepseskaf and Userkaf, of the IV and V Dynasties. Here we continue discussing their burial places, that is their pyramids, in a new framework based on the study of orientation of pyramid substructures (entrances, corridors, burial chambers). Shepseskaf and Userkaf opted for the same architectonic solutions, exhibiting a continuity in funerary architecture. After showing the substructure orientations of the pyramids of the IV and V Dynasties, we will stress that the substructure in the Shepseskaf's pyramid, the Mastabat Fara'un, is oriented as in the Fourth Dynasty architecture of pyramids and that it has a layout which is closer to that used by the following Fifth Dynasty. In our discussion, we will start with the first ruler of the Fourth Dynasty, Snefru, who introduced a new form of external and internal layout of the burial complexes of the king, through an evolution based on four attempts, the Seila, Meidum, Bent and Red pyramids. In the final model, the Snefru's Red pyramid, the substructure has a north entrance and a burial chamber east-west oriented. This orientation persisted in the pyramids of the Fourth and Fifth Dynasties, and beyond. The Sekhemkhet (Djoserty) pyramid of the Third Dynasty is also illustrated for comparison, such as some earlier burial monuments. In the discussion here proposed the position of sarcophagus and of the body inside it is also investigated. The sarcophagus has its axis north-south. The body of the deceased king was lying on its left side, extended, head to the north, face towards the east. In this framework, we propose again what we told in arXiv 2016, arxiv:1604.05963, and the layout of pyramids with respect to sunrise on solstices.

**Introduction**

In arXiv, https://arxiv.org/abs/2412.20407, the subjects of my discussion were the Egyptian kings Shepseskaf and Userkaf, of the IV and V Dynasty respectively, in relation to the eclipse of 2471 BC, a total eclipse visible from the Delta of the Nile. In this previous arXiv, chronology was the main concern. Here, the theme regards the burial places of these two kings, in the framework of the architecture of pyramids. We propose a new approach based on the study of orientation of pyramid substructures (entrances, corridors, burial chambers, sarcophagi). Therefore, after an investigation of the orientations of the pyramids of the IV and V Dynasties, we will stress that the substructure of Shepseskaf's pyramid, known today as Mastabat Fara'un, is coherently positioned in the Fourth Dynasty architecture of pyramids. Moreover, the substructure has a layout which is like that of following Fifth Dynasty.

Our discussion starts with the first ruler of the Fourth Dynasty, Snefru, who introduced a new form of external and internal layout of the burial places of the king, through an evolution based on four attempts, the Seila, Meidum, Bent and Red pyramids. In the final model, the Snefru's Red pyramid, the substructure has a north entrance and a burial chamber east-west oriented. This orientation persisted in the pyramids of the Fourth and Fifth Dynasties and beyond. The Sekhemkhet (Djoserty) pyramid of the Third Dynasty is also illustrated for comparison, such as some earlier burial monuments.

The orientation of the burial chamber is fundamental for the position of the sarcophagus, close to the west end of the chamber, to have a sarcophagus orientation along the north-south axis. The body inside was oriented head to the north and feet to the south, laying on its left side, extended, face towards east. To me, it seems that the body was positioned like the Nile, flowing from south to north. In this framework, we propose again what told in 2016 arXiv, https://arxiv.org/abs/1604.05963 , about the layout of pyramids with respect to sunrise on solstices.

Before starting our discussion, let us stress that the subdivision of Egyptian rulers in Dynasties had origin during the Ptolemaic period, whereas the ancient Egyptians perceived the kingship as a continuous, unbroken sequence of kings. This continuity is evidenced by the fact that Shepseskaf and Userkaf opted for the same architectonic solution, evidencing a continuity in funerary architecture. Orientation and layout of Shepseskaf's substructure of the monument persisted in the pyramids of the Fourth and Fifth Dynasties and beyond.

**From the first Pharaoh to Ramesses II**

"The division of ancient Egyptian kings into dynasties is an invention of Manetho's Aegyptiaca, intended to adhere more closely to the expectations of Manetho's patrons, the Greek rulers of Ptolemaic Egypt." (https://en.wikipedia.org/wiki/Shepseskaf, mentioning Redford, 2001). "In modern Egyptology no sharp division is understood to have taken place between the fourth and fifth dynasties" (Wikipedia, Bárta, 2016). And "We must remember that the whole concept of dynasties is a later construct, that the Egyptians themselves saw their kings in an unbroken line" (Hawass, 2006). This 'unbroken line' is also evidenced by the fact that the Shepseskaf' burial place, the Mastabat Fara'un, was restored by Khaemwaset, the fourth son of Ramesses II. And Khaemwaset restored also the Userkaf pyramid. We know that Mastabat Fara'un is the burial place of Shepseskaf thanks to Khaemwaset's inscription on it. Due to the interest of Khaemwaset for the monuments of the Old Kingdom, this son of Ramesses II is considered the first 'Egyptologist' of history.

In the Khaemwaset inscription, we can find that the Shepseskaf's burial place was defined as a pyramid. We can see also the hieroglyph of the "pyramid" on a "fragment of limestone relief from the tomb of Shepseskaf: this slab bears a single line of text, carved in fair sunk relief, containing the name of the 'Pyramid' of King Shepseskaf", at the web page of the British Museum: https://www.britishmuseum.org/collection/object/Y_EA1234.

Regarding Khaemwaset, we must refer to an article by Lloyd, 2024. In the Past article, https://the-past.com/feature/the-first-egyptologists/, entitled "The First Egyptologists", we find that Khaemwaset "enjoyed a brilliant career across the entire spectrum of activities expected of a royal prince, but is particularly renowned for his religious and architectural activities. We are fortunate in having a number of inscriptions that give an insight into his motivation for working on ancient monuments, and of these texts I [Lloyd] want to examine two: the inscription on the Mastabat Fara'un constructed for the Fourth Dynasty king Shepseskaf (c.2503-2498 BC) at Saqqara South, and the text recording Khaemwaset's work on the statue of Kawab, eldest son of Khufu" (Lloyd, 2024).

The inscription of the Shepseskaf mastaba tells to us: "His Majesty instructed the Chief Controller of Craftsmen [i.e. High Priest of Ptah], Sem-Priest, King's Son, Khaemwaset to establish the name of the King of Upper and Lower Egypt Shepseskaf [i.e. on the monument], since his name could not be found on his pyramid, inasmuch as the Sem-Priest Khaemwaset loved to restore the monuments of the Kings of Upper and Lower Egypt, because of what they had achieved, making firm again what had fallen into ruin. By [the Sem-Priest, King's Son, Khaemwaset]" (Lloyd, 2024).

Look at the inscription, so you can see that there is the symbol of the pyramid.

https://i0.wp.com/the-past.com/wp-content/uploads/2024/12/post-1_image2-14.jpg?w=1230&ssl=1

For Khaemwaset, Shepseskaf's mastaba was a pyramid.

Khaemwaset is acting at Mastabat Fara'un "on his father's command". Lloyd stresses that the inscription is related to the "concerns over legitimacy, which was of particular importance to this family because their right to the throne of Egypt was far from obvious. This preoccupation is particularly evident in the cenotaph of Sety I at Abydos. There we find a spectacular king-list depicting Sety making offerings to a long sequence of kings beginning with the First Dynasty, and running right down to the Nineteenth, a list that carefully avoids mentioning any ruler (such as Akhenaten) who might be regarded as illegitimate" (Lloyd, 2024). Therefore, it was not a problem of 'Dynasty', but a problem of legitimacy. For what is regarding Ramesses' Dynasty, "It was possible to assert [its] legitimacy by showing concern for the monuments of predecessors" (Lloyd, 2024).

"Khaemwaset restored the monuments of earlier kings and nobles. Restoration texts were found associated with the pyramid of Unas at Saqqara, the tomb of Shepseskaf called the Mastabat al-Fir'aun, the sun-Temple of Nyuserre Ini, the Pyramid of Sahure, the Pyramid of Djoser, and the Pyramid of Userkaf. Inscriptions at the pyramid temple of Userkaf show Khaemwaset with offering bearers, and at the pyramid temple of Sahure Khaemwaset offers a statue of the goddess Bast." https://ancientegypt.fandom.com/wiki/Khaemwaset_A#Restoration_projects

"Khaemwaset is attested at the sun temple of Niuserre" (Almansa-Villatoro et al., 2022, mentioning Gomaà, 1973). "Distinctive markers of Khaemwaset's activities are the inscriptions in the names of he and his father added to several standing monuments in the Memphite necropolis and at other sites. These texts are known both from surviving examples but may in some cases be attested indirectly from later accounts of the appearance of monuments. The Unas pyramid inscription was found on its south side [41], and is now restored in that position." (Price, 2022, mentioning Drioton and Lauer, 1937).

"It is the best preserved of such texts and is typical in formulation, being framed by a royal command: 'His Majesty decreed an announcement; It is the High Priest (of Ptah), the Sem-priest, King's Son Khaemwaset, who has perpetuated the name of King [Unas]. Now his name was not found on the face of his pyramid. Very greatly did the Sem-priest, King's Son Khaemwaset, desire to restore the monuments of the Kings of Upper and Lower Egypt, because of what they had done, the strength of which was falling into decay. He set forth a decree for its sacred offerings,… its water … [endowed] with a grant of land, together with its personnel…'. (writer's translation)" (Price, 2022).

"Other such inscriptions appear in different positions on individual monuments: on the north side of the mastaba of Shepseskaf [1] (p. 77, no. 12); the south side of the Step Pyramid [1] (p. 77, no. 8); the east side of the pyramid of Userkaf [1] (p. 77, no. 9); the south side of the sun temple of Niuserre at Abu Ghurab [1] (p. 76, no. 4); an unknown location on the pyramid of Sahure at Abusir [41] (pp. 205–207); and on the south side of the pyramid of Pepi I at Saqqara [42]." (Price, 2022, referring to [1] Gomaà, [41] Drioton and Lauer, 1937, and [42] Leclant, 1993).

**The layout of the pyramids**

After this ample preamble on the fact that ancient Egyptians did not perceive the notion of 'Dynasty' and that the son of Ramesses II took care of restoring monuments of those who were considered as his 'ancestors', let us pass to the study of the substructure of pyramid. That is, to understand Shepseskaf' Mastabat, we need to understand the pyramids. We will start from the Fourt Dynasty and continue with the Fifth Dynasty. That is, we continue to use Manetho's conventional subdivision in Dynasties.

Snefru, who was the first king of the IV Dynasty of the Old Kingdom, built four pyramids. Let us consider Muhlestein, 2019, and his Table 1, but before showing the Table, let us report Owen Jarus, 2009, and some of their features. After showing "all four of Snefru's pyramids are like, Muhlestein is able to offer a comparison. He found some interesting parallels between the pyramids. These include – 1) Both Seila and Meidum were built at almost the exact same angle. Meidum is 51 degrees, Seila is 52. On the other hand, both the Bent and the Red Pyramids have an angle of 43 degrees. – 2) Both Seila and Meidum have causeways that lead nowhere – there is no building at the end of them. – 3) The Red and Bent pyramids have causeways that lead to buildings. - 4) The Bent pyramid and the Red Pyramid have Valley Temples. Meidum and Seila have no evidence of any Valley Temples. - 5) Meidum and the Red Pyramid have mortuary temples. The Seila and Bent the pyramid do not." (Jarus, 2009). "All four pyramids appear to have altars." (Jarus, 2009).

This is the evolution of the outer layout of the pyramids. Let us see the above-mentioned pyramids in satellite Google Maps.

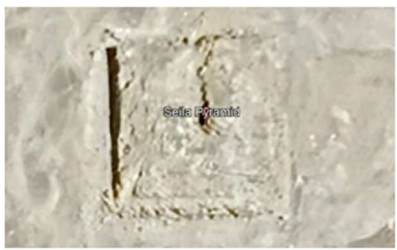 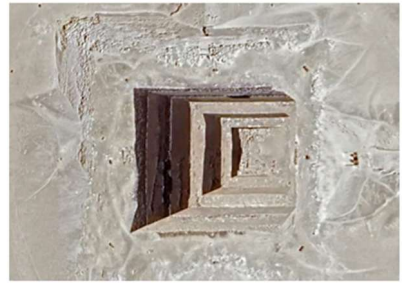

Above. Seila Pyramid (left), Courtesy Google Earth. The size of the pyramid, as is visible in the image, is about 25 m. Meidum pyramid (right), Courtesy Google Earth. The side of the pyramid, as visible in the image, is about 64 m. Below. The Bent Pyramid (left), Courtesy Google Earth. Side 190 m. The Red Pyramid (right), Courtesy Google Earth. Side 208 m.

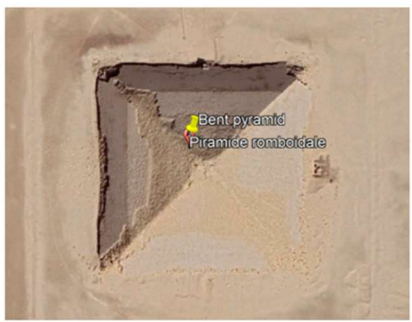 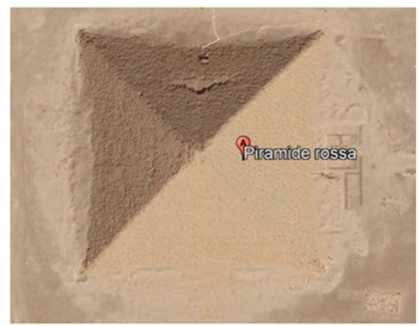

**Snefru and his pyramid substructures**

Let us consider the evolution inside.

Table I, Muhlestein, 2019, is giving information about the burial chambers.

| Pyramid | Burial Chamber Orientation | Side of Entrance | Adjacent Structure | Causeway | Valley Temple | Altars |
|---|---|---|---|---|---|---|
| Seila | None (uncertain) | None (uncertain) | Porch on East and North | East | No | Two on North, perhaps one on East |
| Meidum | North-South | North | East temple | East | No | One on East |
| Bent | North-South | North and West | East and North temple | North-East, largely East | Yes | One on East, one on North |
| Red | East-West | North | East temple | East | Yes | One on East |

Note the evolution of the orientation of the burial chamber. http://www.narmer.pl/pir/snofru_en.htm

Meidum: "The burial chamber measures 5.9 by 2.65 metres, which is quite small, yet another sign that the builders were experimenting. There is no sarcophagus and no trace of a burial. Outside the pyramid many elements that would become the standard for pyramid complexes to come were already present as well."

https://www.memphistours.com/Egypt/WikiTravel/Pyramids-Egypt/wiki/Pyramid-of-Meidum .

Bent Pyramid: "The design of the corridors is similar to the one found in the Great Pyramid of Giza, where the Grand Gallery takes up the place of the ascending corridor. The corridor leads up to the burial chamber (called this despite that it most probably never contained any sarcophagus)." (https://en.wikipedia.org/wiki/Bent_Pyramid, mentioning Fakhry, 1961).

"Archaeologists have discovered a large structure to the northeast of the 4,600 year old Bent Pyramid which may be the remains of an ancient harbour. It connects to one of the pyramids temples by way of a 140 meter long causeway." (Jarus, 2010). "Although the Red Pyramid is often associated with Pharaoh Snefru, no direct evidence of his burial or funerary items has been found within it." https://www.encounterstravel.com/eu/blog/red-pyramid.

**The Bent Pyramid**

"The pyramid is exceptional in that it has two largely separate internal layouts; a lower one with an entrance located on the north face at a height of 11.33 meters from the ground level, and an upper one – a unique case for the Old Kingdom – with an entrance situated on the west face at a height of 32.76 meters. These two systems of apartments both contain a vast chamber which is covered with a corbelled vault. These different arrangements were connected by a gallery dug through the existing masonry, undoubtedly by the builders themselves at a later stage of the construction work, but before the western access was definitively closed. The premature blocking of the western descending passage, which then forced the builders to reach the upper chamber from the lower system, has found no explanation so far" (Monnier & Puchkov, 2016). See the substructures inside the pyramid in the Monnier and Puchkow article.

**The Red Pyramid**

"Visitors climb steps cut in or built over the stones of the pyramid to an entrance high on the north side. A passage, 3 feet (0.91 m) in height and 4 feet (1.2 m) wide, slopes down at 27° for 200 feet (61

m) to a short horizontal passage leading into a chamber whose corbelled roof is 40 feet (12 m) high and rises in eleven steps. At the southern end of the chamber, but offset to the west, another short horizontal passage leads into the second chamber. This passage was probably closed at one time and the offset was a measure intended to confuse potential tomb robbers. The second chamber is similar to the first and lies directly beneath the apex of the pyramid. High in the southern wall of the chamber is an entrance, now reached by a large wooden staircase built for the convenience of tourists. This gives onto a short horizontal passage that leads to the third and final chamber with a corbelled roof 50 feet (15 m) high. The first two chambers have their long axis aligned north-south, but this chamber's long axis is aligned east-west. Unlike the first two chambers, which have fine smooth floors on the same level as the passages, the floor of the third chamber is very rough and sunk below the level of the access passage. It is believed that this is the work of tomb robbers searching for treasure in what is thought to have been the burial chamber of the pyramid". https://en.wikipedia.org/wiki/Red_Pyramid . To see the pyramid inside, please use the following link https://www.guardians.net/egypt/red2.htm

**IV and V Pyramids**

Let us add the three well-known Giza Pyramids and other pyramids and Mastabat Fara'un, in the following Table for the Fourth Dynasty. In the same Table, the substructure orientations of the Fifth Dynasty are given too.

| Pyramid (IV Dynasty) | Burial Chamber Orientation | Side of Entrance |
|---|---|---|
| Khufu | Chamber East-West, Sarcophagus North-South (see link, https://pathamid.com/path07.html ) | North |
| Queen Hetepheres | Chamber East-West? | North |
| Djedefre | Chamber East-West . See the map at (*) Sarcophagus North-South, as shown by the picture https://www.flickr.com/photos/manna4u/6323876246 | North |
| Khafre | Chamber East-West, Sarcophagus North-South https://guardians.net/egypt/pyramids/Khafre/KhafrePyramid.htm | North |
| Bicheris | Chamber East-West . See the map at (**) | North |
| Menkaure | Antechamber East-West, Sarcophagus North-South http://www.narmer.pl/pir/menkaure_en.htm | North |
| Shepseskaf | Chamber East-West, The sarcophagus is broken | North |
| (V Dynasty) | | |
| Userkaf | Chamber East-West, The sarcophagus is broken | North |
| Sahure | Chamber East-West (damaged) | North |
| Neferirkare Kakai | Chamber East-West | North |
| Neferefre Isi | Chamber East-West | North |
| Nyuserre | Chamber East-West | North |
| Djedkare Isesi | Chamber East-Wets Sarcophagus North-South | North |
| Unas | Chamber East-West, Sarcophagus North-South https://www.sofiatopia.org/maat/iunas12.jpg | North |

(*) https://www.ancient-egypt.org/history/old-kingdom/4th-dynasty/djedefre/pyramid-complex-at-abu.html

(**) https://en.wikipedia.org/wiki/Unfinished_Northern_Pyramid_of_Zawyet_El_Aryan#/media/File:Baka-Pyramide_Plan.png

**Khufu Pyramid**. https://pathamid.com/path07.html  "At the end of the second low passage we come to the last chamber of the Pyramid, a large room made out of granite blocks that has been subjected to damage from earthquake or settlement during its past. It houses a coffer made from the same enduring granite, without lid and of rough finish. The coffer is believed to have been originally sited centrally lengthwise along the north-south axis of the pyramid, bisected by the east-west axis of the chamber and for the following exploration this will be assumed as its correct position." (Edwards, 2011).

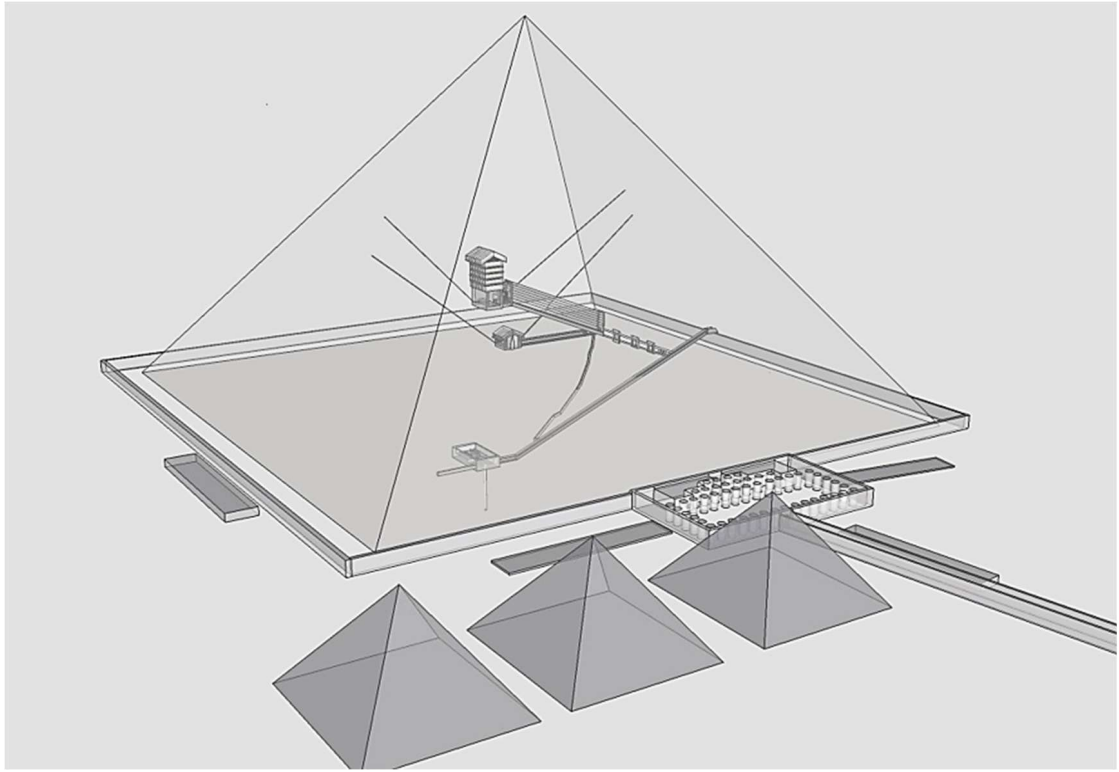

Original image by R.F.Morgan. Uploaded by Mark Cartwright, published on 19 December 2016. The copyright holder has published this content under the following license: Creative Commons Attribution-ShareAlike.

https://www.worldhistory.org/image/6194/interior-design-great-pyramid-of-giza/

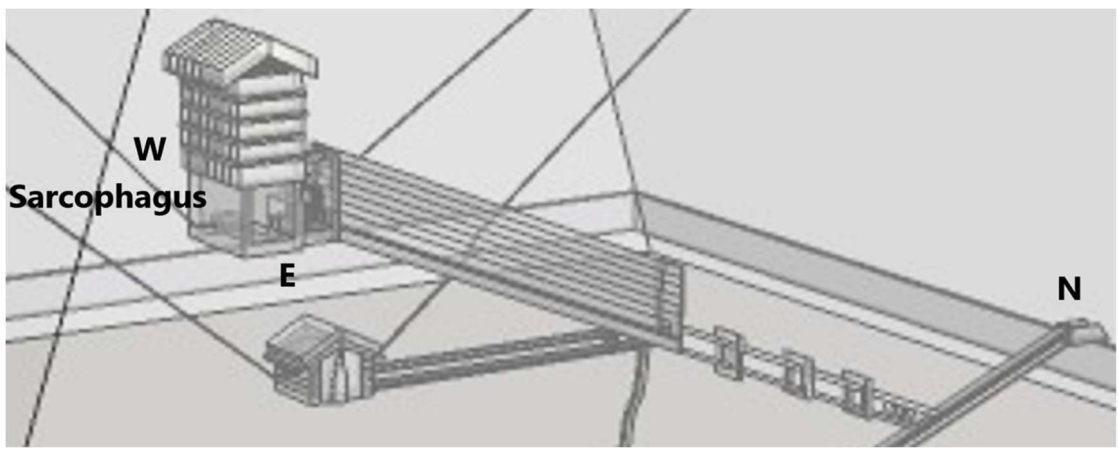

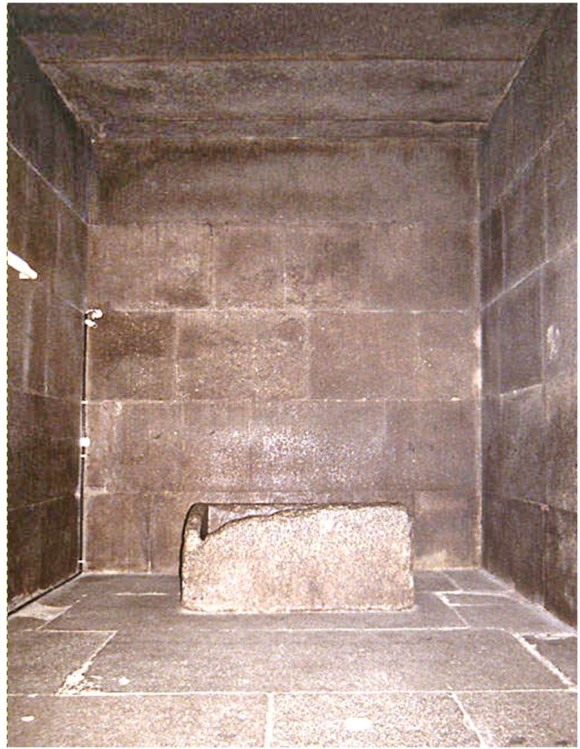

Sarcophagus in the King's Chamber. Courtesy Jon Bodsworth - https://commons.wikimedia.org/wiki/File:Chambre-roi-grande-pyramide.jpg?uselang=en#Licensing

**G1a (Queen Hetepheres)**

"The northern-most pyramid (G1a) was originally ascribed to Queen Meritetes (or Mertitiotes), but is now considered to be the secondary burial of Queen Hetepheres I . … The entrance is in the north wall, just off the north-south axis. Inside the pyramid, a corridor descends to the mid-point of the structure before turning right into a small burial chamber cut into the rock and surfaced with limestone blocks. Although there is a recess carved into the west wall of the chamber, no sarcophagus was found in the tomb." https://ancientegyptonline.co.uk/queens-pyramids/

**Djedefre (Radjedef) pyramid**. It "was architecturally different from those of his immediate predecessors in that the chambers were beneath the pyramid instead of inside. The pyramid was built over a natural mound and the chambers were created using the "pit and ramp" method, previously used on some mastaba tombs. Djedefre dug a pit 21m x 9m and 20m deep in the natural mound. A ramp was created at an angle of 22º35' and the chambers and access passage were built within the pit and on the ramp. Once the 'inner chambers' were finished, the pit and ramp were filled in and the pyramid built over the top. This allowed the chambers to be made without tunneling, and avoided the structural complications of making chambers within the body of the pyramid itself. He also reverted to an earlier style of construction by creating a rectangular enclosure wall oriented north-south, similar to those of Djoser and Sekhemkhet" https://en.wikipedia.org/wiki/Pyramid_of_Djedefre

"The royal pyramid stood almost in the centre of the complex. The pyramids of Djedefre's predecessors Snofru and Kheops had the burial chamber inside the pyramid above ground level. For unknown reasons, Djedefre prefered to have his burial chamber, built at the bottom of a collossal pit measuring 23 by 10 metres and sunk some 20 metres into the ground. The burial chamber itself measured 21 by 9 metres. This technique was also used for the building of the burial chamber of Netjerikhet at Saqqara. A 49 metre long corridor slopes up to ground level, providing the entrance to the pyramid. As was already traditional, this entrance was located in the north, pointing to the circumpolar stars." https://www.ancient-egypt.org/history/old-kingdom/

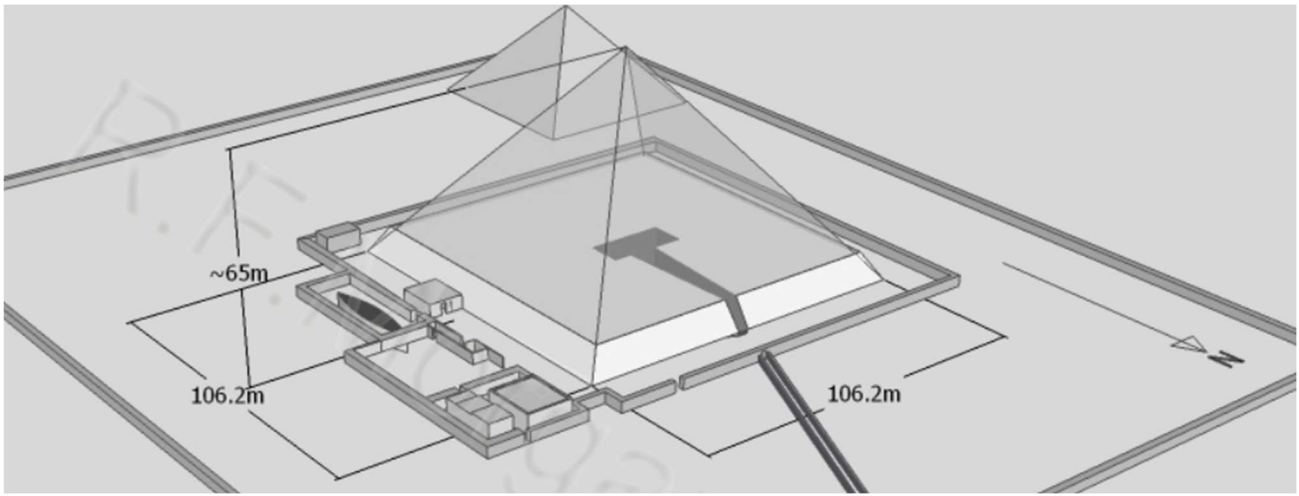

The pyramid complex of Djedefre, Courtesy, R.F. Morgan,
https://commons.wikimedia.org/wiki/File:008_Djedefre1.jpg

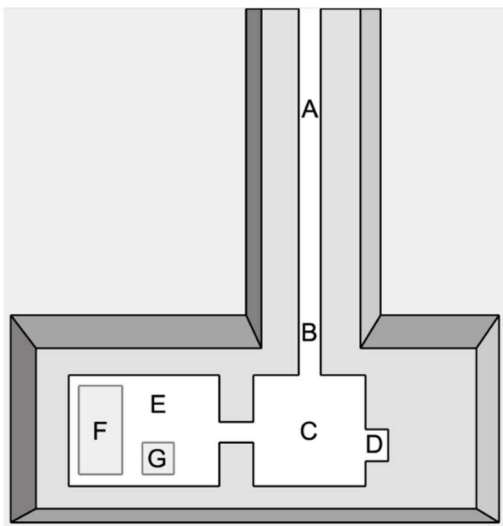

Reconstruction of the substructure within the excavation of the Djedefre (Radjedef) pyramid: A = descending passage B = horizontal passage C = antechamber D = niche (or serdab?) E = burial chamber F = sarcophagus G = canopic chest. Courtesy GDK,

https://de.wikipedia.org/wiki/Radjedef-Pyramide#/media/Datei:Radjedef-Pyramide_Substruktur.png

"The bottom of the chamber was bricked up with five layers of fine limestone to reach the level of the horizontal passage. On the limestone floor, only small remains of the chambers could be detected. Apparently, there was an antechamber under the center of the pyramid, from which a passage led to the west to the burial chamber. The antechamber probably had a niche or an extension to a serdab on the east side. [21] The floor of the burial chamber has depressions that indicate that a sarcophagus and a canopic chest were embedded here, similar to the pyramid of Khafre. [21] In the pit it was found a fragment of a large granite beam, one end of which did not end at an angle of 90° to the side surface, but at 135°. From this it can be concluded that this was part of the gable roof of the burial chamber.

There is also an inscription on this fragment that refers to Radjedef. [19] A limestone block with similar inscriptions was also found. [18]" (Wikipedia, mentioning [18] Valloggia, 2001, [19] Grimal, 1997, [21] Grimal, 1998).

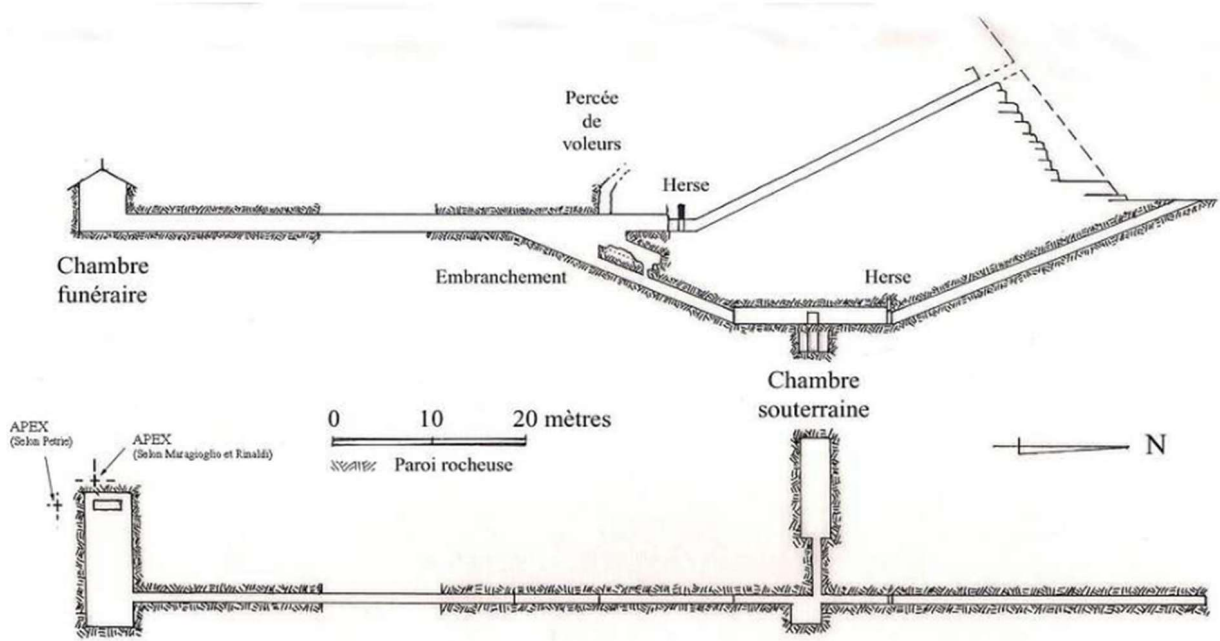

Map of the funerary apartments of the pyramid of Chephren. Plan des appartements funéraires de la pyramide de Khephren. 2007. Courtesy Franck Monnier d'après John Shae Perring (operations carried on at pyramids of Gizeh) et Maragioglio et Rinaldi.
https://commons.wikimedia.org/wiki/File:Appartements-funéraires-khephren.jpg

**Khafre Pyramid**.  https://guardians.net/egypt/pyramids/Khafre/KhafrePyramid.htm  "Continuing south down the passageway leads to the main burial chamber. On this higher level there is a chamber which is ft. 46.5 ft. long and 16.5 ft. wide. The ceiling also comes to a point. There is a unique black granite sarcophagus in this room in that it was built to be sunken into the floor. The original lid, though no longer attached, lies propped up next to the coffer near the west wall. It is possible that the open niche against the east side of the coffer held the king's canopic chest, the box containing the mummified organs of the king, within ceremonial vases. There are a few other examples of this style in other Old Kingdom tombs".

**Bikheris (Baka) Pyramid.** "The Unfinished Northern Pyramid of Zawyet El Aryan, also known as Pyramid of Baka and Pyramid of Bikheris is the term archaeologists and Egyptologists use to describe a large shaft part of an unfinished pyramid at Zawyet El Aryan in Egypt. Archaeologists are generally of the opinion that it belongs to the early or the mid-4th Dynasty (2613–2494 BC) during the Old Kingdom period. The pyramid owner is not known for certain and most Egyptologists, such as Miroslav Verner, think it should be a king known under his hellenized name, Bikheris, perhaps from the Egyptian Baka.[1] In contrast, Wolfgang Helck and other Egyptologists doubt this attribution.[2]" https://en.wikipedia.org/wiki/Unfinished_Northern_Pyramid_of_Zawyet_El_Aryan . [1] is Verner, 1999, and [2] Helck and Otto, 1984.

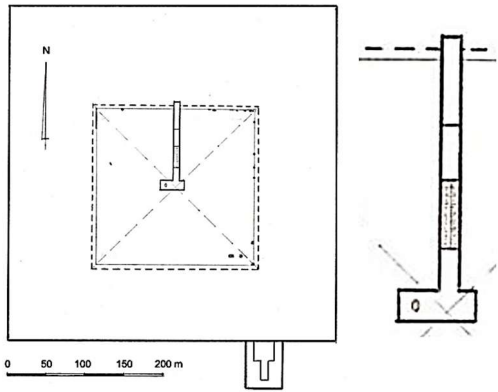

"Floor plan of the pyramid of Baka. Plan-complexe-grande-excavation.jpg: MONNIER Franck / derivative work: GDK (talk) - Plan-complexe-grande-excavation.jpg Complexe funéraire de la grande excavation de zaouiet el-aryan" Available e.nwikipedia

https://en.wikipedia.org/wiki/Unfinished_Northern_Pyramid_of_Zawyet_El_Aryan#/media/File:Baka-Pyramide_Plan.png

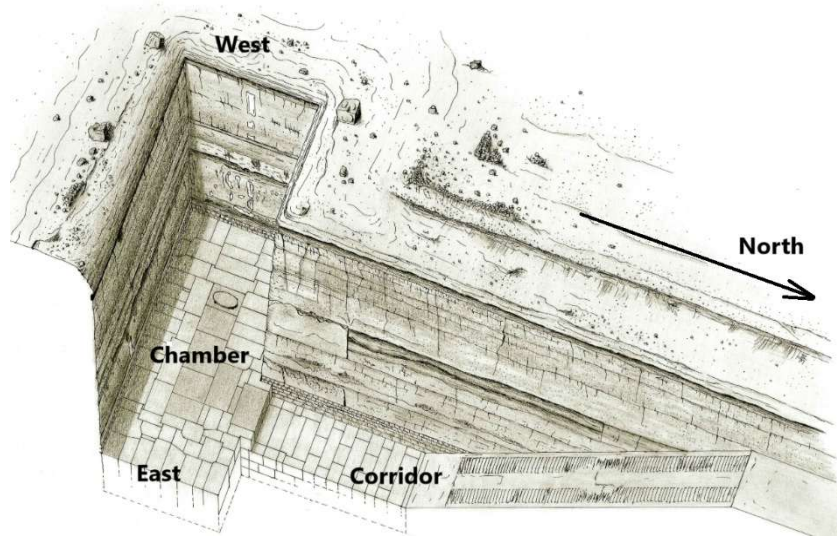

Excavated Corridor and Burial Chamber at Zaouiet el-aryan. Courtesy MONNIER Franck,
https://en.wikipedia.org/wiki/Unfinished_Northern_Pyramid_of_Zawyet_El_Aryan#/media/File:Vue-grande-excavation.jpg

**Menkaure pyramid**: http://www.narmer.pl/pir/menkaure_en.htm "On the northern side, from the low situated entrance at a height of 4 m, a descending corridor with a length of 32 m leads to a chamber with a niche decoration, a granite chamber with three stone barriers-traps. After almost 13 m there is a vestibule chamber. … In front of the funerary chamber there are stairs, leading to a room with six niches, possibly serving as warehouses. Menkaure's sarcophagus decorated with the image of the palace facade, removed from the pyramid in 1837 by R.W.H. Vyse, sank along with a transport ship off the coast of Europe in the Bay of Biscay. The remains of a wooden coffin from the vestibule, most probably from the time of the Saitian renewal of the burial site, found themselves in London. The mummy bones and bandages found at that time probably come from early Christianity."

https://guardians.net/egypt/pyramids/Menkaure/MenkaurePyramid.htm "Down the passageway from the previous chamber leads to the main burial chamber. There is a finely finished interior of the vaulted ceiling. The niche in the floor housed the original sarcophagus. This sarcophagus was removed from the pyramid and shipped on a boat to England on Oct. 1838. The boat sank on the way, and the sarcophagus of Menkaure has not been seen since."

**The Decree**

Strudwick, 2005. Proposes the "Texts from the pyramid age", and we can find a decree regarding Menkaure and Shepseskaf.

"This is the earliest decree to survive from the Old Kingdom. Dated to the first year of Shepseskafs reign, it suggests that it was set up at least partly as part of his pious completion of the earlier temple—it would appear that Menkaure died before the complex was complete and that Shepseskaf finished off the temple in brick rather than stone for the sake of speed (Reisner 1931: 29-31). It is just possible that this decree is a later copy of an original document, suggested by the prominence of the cartouche name at the top of the decree, which is rather strange for the Old Kingdom. However, its rather unusual structure perhaps argues for an early date for the original, before the more established forms had evolved. The fragments are now in the Egyptian Museum in Cairo (Temp. no. 26.2.21.18)."

And here the Decree as given by Strudwick: "Horus Shepseskhet, the year after the first occasion of the count of the cattle and herds ... which was done in the presence of the king himself. The king of Upper and Lower Egypt Shepseskaf. For the king of Upper and Lower Egypt [Menkaure] he set up a monument (in the form of) a pekher offering ... in the pyramid of Menkaure ... With regard to the pekher offering brought for the king of Upper and Lower Egypt [Menkaure] ... priestly duty [is done] with respect to it for ever … in the course of his duty for ever … the pyramid of Menkaure … burial … the pyramid of Menkaure …."

**Sekhemkare, from the Fourth to the Fifth Dynasty**

"The tomb of Sekhemkare is known as G 8154 (= LG 89), located in the Central Field which is part of the Giza Necropolis. His tomb is located to the southwest of that of his brother Nikaure and to the northwest of the tombs of Niuserre and Niankhre. The rock-cut tomb is situated to the northwest of the complex of Queen Khentkaus I and southeast of the Pyramid of Khafre. On the doorjambs at the tomb entrance Sekhemkare is identified as "the king's son of his body, hereditary prince, count, councilor, sealer of the king of Lower Egypt, chief lector-priest of his father, sole companion, secretary of the House of Morning, assistant of (the god) Duau, secretary of his father". In other scenes he has the additional titles of chief justice and vizier and director of the palace. In one of the rooms an inscription records his career at court during the reigns of successive kings" https://en.wikipedia.org/wiki/Sekhemkare_(vizier)

We can find all the details in Hassan, 1943, regarding Excavations at Giza: IV: 1932-33.

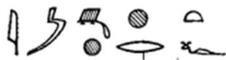
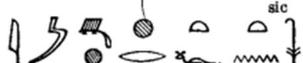
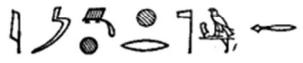
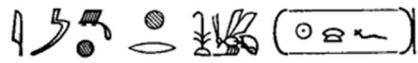

| | | |
|---|---|---|
| 21. | *imꜣḫw ḫr njswt-bjtj Mn-kꜣw-Rꜥ.* | Honoured by the King of Upper and Lower Egypt, Men-kaw-Raʿ. |
| 22. | *imꜣḫw ḫr njswt-bjtj Špss-kꜣ-f.* | Honoured by the King of Upper and Lower Egypt, Shepses-ka-f. |
| 23. | *imꜣḫw ḫr njswt-bjtj Wsr-kꜣ-f.* | Honoured by the King of Upper and Lower Egypt, Weser-ka-f. |
| 24. | *imꜣḫw ḫr njswt-bjtj Sꜣḥw-Rꜥ.* | Honoured by the King of Upper and Lower Egypt, Saḥw-Raʿ (¹). |

Before the Shepseskaf' pyramid, let us mention that of Userkaf.

**Userkaf Pyramid**: "The pyramid's entrance is located in the center of its north side and opens onto a substructure that is entirely underground. An 18.5 meter long descending passage goes down to a horizontal corridor, that was partially clad with granite blocks and in the middle of which was a huge portcullis slab. Almost immediately behind the portcullis, a short corridor to the east opened on a T-shaped magazine. Further down the main corridor was an antechamber of 4.14 by 3.12 metres. To the west, this antechamber opened onto the actual burial chamber, that measured 7.87 by 3.13 metres (see cut-away of pyramid). The burial chamber was originally completely lined and paved with fine limestone. … The basalt sarcophagus was found empty", 5th-dynasty/userkaf/pyramid-complex-at-saqqara/pyramid-of-userkaf.html . The offering chapel is to the east of the pyramid.

"The interior of the pyramid of Userkaf at Saqqara begins with an entrance passageway blocked by a granite portcullis which has been broken through. Beyond this a couple of storage rooms - where we enjoyed dodging the bats! - lie off to the left, while ahead there's an antechamber and then the burial chamber itself. The latter contains only the smashed fragments of a basalt sarcophagus, but it's an impressive space nonetheless." (Chris Naunton,

https://www.facebook.com/chrisnauntonofficial/posts/the-interior-of-the-pyramid-of-userkaf-at-saqqara-begins-with-an-entrance-passag/1111105037127274/).

**Sahure Pyramid:** "The pyramid of Sahura was first investigated by John Perring, who was the only one who was able to break off and clean the entrance as well as the descending access passage. The burial chamber was very badly damaged by the stonecutters, and it is not even clear if it is comprised of one or two rooms. Perring found a single fragment of basalt, thinking that it belonged to the king's sarcophagus. Interestingly, in the northeastern part of the eastern wall of the burial chamber, Perring discovered a low passageway, indicated as "C" on his plan. He suggested that this could lead to a magazine area, however, the corridor was full of rubble or waste and he did not attempt to entre it. Due to the bad state of preservation in the interior compartment of the pyramid, precise reconstruction of the substructure's plan was impossible."

https://arce.org/project/pyramid-complex-king-sahura-protection-restoration-and-documentation/

**Neferirkare Kakai Pyramid**: "The descending corridor near the middle of the north face of the pyramid serves as the entry into the substructure of Neferirkare's pyramid. … The corridor breaks out into a vestibule leading to a longer corridor which is guarded by a portcullis. This second corridor has two turns, but maintains a generally eastward direction and ends in an antechamber offset from the burial chamber. … The burial and ante chamber's ceilings were … Thieves have ransacked the chambers of its limestone making it impossible to properly reconstruct, though some details can still be discerned. Namely, that (1) both rooms were oriented along an east–west axis, (2) both chambers were the same width; the antechamber was shorter of the two, and (3) both chambers had the same style roof, and are missing one layer of limestone. Overall, the substructure is badly damaged: the collapse of a layer of the limestone beams has covered the burial chamber. No trace of the mummy, sarcophagus, or any burial equipment has been found inside. The severity of the damage to the substructure prevents further excavation." https://en.wikipedia.org/wiki/Pyramid_of_Neferirkare

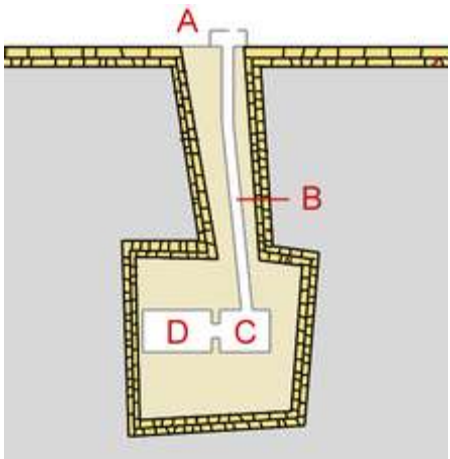

**Neferefre Pyramid**: "The pyramid substructure was accessed from slightly above ground level on the middle of the pyramid's north side. A descending corridor, deflected slightly to the south-east, led to the funerary apartments. … The corridor terminates at an antechamber, with a burial chamber lying further to the west. The rooms are oriented along the east-west axis and each apartment was originally covered by a gabled fine white limestone ceiling. … In spite of the devastation wrought by stone thieves, remnants of the burial have been preserved. Inside the substructure fragments of a red granite sarcophagus, pieces of four alabaster canopic jars, alabaster sacrificial offering containers, and a partial mummy were recovered". https://en.wikipedia.org/wiki/Pyramid_of_Neferefre The plan is a Courtesy https://commons.wikimedia.org/wiki/File:Raneferef-Pyramide_Substruktur.png

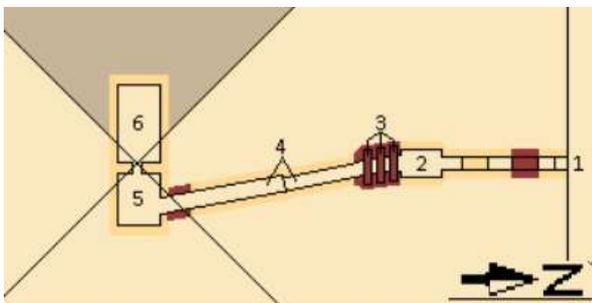

**Nyuserre Pyramid:** "Layout of Nyuserre's substructure. In order: (1) North facing entry; (2) Vestibule; (3) Pink granite portcullises; (4) Passageway; (5) Antechamber; (6) Burial chamber. Granite presented in red, limestone presented in orange." Image courtesy of Mr rnddude, https://en.wikipedia.org/wiki/Pyramid_of_Nyuserre

Reference: Borchardt, Ludwig (1907) Das Grabdenkmal des Königs Ne-User-Re, Leipzig: Hinrichs, p. Blatt 19.

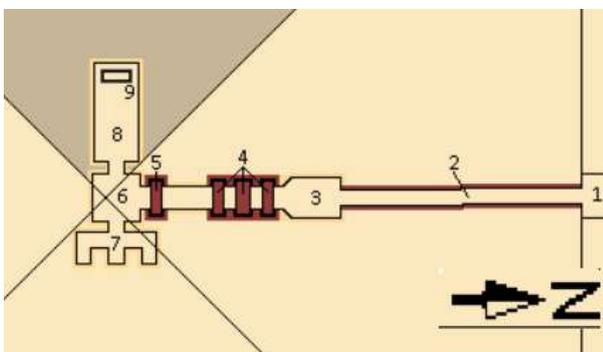

**Djedkare Isesi Pyramid:** Layout of Djedkare Isesi pyramid substructure. 1) North chapel; 2) Descending corridor; 3) Vestibule; 4 and 5) Granite portculli; 6) Antechamber; 7) Serdab; 8) Burial chamber; 9) Sarcophagus. Courtesy Mr rnddude, https://en.wikipedia.org/wiki/Pyramid_of_Djedkare_Isesi#/media/File:Djedkare's_Substructure.png

"To the west lay the burial chamber, measuring 7.84 m (25.7 ft) by 3.1 m (10 ft), which once contained the basalt sarcophagus of the ruler. Fragments of the sarcophagus were found in a 13 cm (5 in) depression in the floor." https://en.wikipedia.org/wiki/Pyramid_of_Djedkare_Isesi

**Unas Pyramid:** "Entering the pyramid from the North, it is necessary to bend over in order to move down the passage. The slope is deliberate and varies between 28° (Khufu), 26° (Khafre), 25° (Pepi II) or 22° in the case of the pyramid of Unas. The passage is oriented to specific northern stars. It slopes down to a corridor-chamber or vestibule, followed by the usual horizontal passage with three granite portcullis slabs. …

Plan of the royal tomb underneath the pyramid of Unas https://www.sofiatopia.org/maat/iunas2.jpg.

This entrance/exit corridor then opens into the antechamber, directly under the pyramid's centre axis. … On the ceiling of the tomb, golden, pentagram-like stars were carved in relief on a sky-blue background. … In the East of the antechamber (on the left hand side when entering the tomb), a doorway opens to the undecorated and uninscribed tomb-chapel with three recesses. …

Burial-chamber - pyramid of King Unas. Sarcophagus West, western half of North & South walls in alabaster, https://www.sofiatopia.org/maat/iunas3.jpg.

On the West of the antechamber (at the right hand side when entering the tomb and precisely opposite the Ka-chapel), a passage-way leads to the burial-chamber. This has a black granite sarcophagus at its West end. In its immediate vicinity, there are no texts. Instead, we see a palace-façade design, with reed-mats and a wood-frame enclosure, an iconography derived from the royal mastaba tombs of the First Dynasty." https://www.sofiatopia.org/maat/wenis.htm . The web site is referring to Naydler, 2005, p.164.

"In the West, the place of regeneration, the mummy is in the total darkness of Osiris, allowing it to be reborn, ascending to illumination. The walls around the sarcophagus, on which these designs were carved, are made of polished alabaster, whereas all the other walls of the tomb are in Tura limestone. Alabaster is soft and translucent". https://www.sofiatopia.org/maat/wenis.htm

**The sarcophagus (Fourth Dynasty)**

The first example is the Mindjedef sarcophagus. "Mindjedef, who lived during the mid- to late 4th Dynasty, was buried in a large tomb on the east side of Khufu's pyramid. This red granite sarcophagus was found in the badly disturbed burial chamber, with a dismembered skeleton laid on top of its lid. The outstretched body would have likely been placed directly inside this stone coffer, perhaps laid on his left side and wrapped in linen covered with a layer of plaster in which his limbs and face would have been molded, as was the practice for the elite during this period. The lugs on the end of the sarcophagus lid would have been used to lower it into place". https://www.metmuseum.org/art/collection/search/552235

The second is has the design of a palace. "Found in a royal tomb, this sarcophagus, or monumental coffin, housed the mummy of a prince or his wife. A pattern of niches imitating the designs on palace walls decorates its surface, alluding to the sarcophagus's function as the deceased's final home. Holes drilled in each end of the lid allowed poles or ropes to be inserted in order to carry it and position it on the box." https://www.brooklynmuseum.org/opencollection/objects/3519

For a detailed description and discussion, wee Sousa, 2014.

**Shepseskaf pyramid, the Mastabat Fara'un.**

Let us stress that the orientation of Mastabat Fara'un is the same as the pyramids of the Old Kingdom.

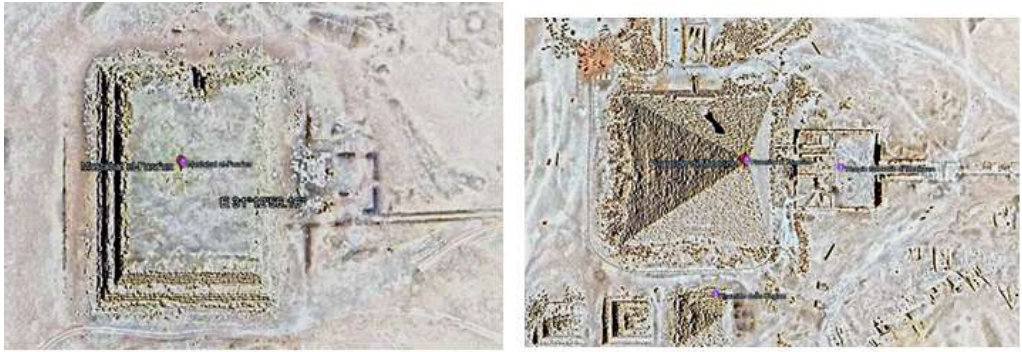

Left. Mastabat Fara'un Courtesy Google Earth. Right. Menkaure's pyramid Courtesy Google Earth

The orientations of the Mastabat Fara'un - entrance, burial chamber - is like those of the Giza Pyramids. The same we find for the Userkaf pyramid. Let us note that the layout of the internal substructures is more similar, for its simplicity, to that of the pyramids of the following dynasties.

"King Shepseskaf … constructed for himself the so-called Mastabat Fara'un, half-way between Saqqara and Dahshur. The form of this tomb differs from the pyramids of the other kings of the Fourth Dynasty. It was a rectangular mastaba construction with a rounded top and vertical end-pieces which gave it the form of the **usual stone sarcophagus**. Inside, the burial apartments were lined with granites. **The heavy masonry and sound workmanship betoken work in the best Fourth Dynasty traditions**" (Cambridge Ancient History). "The building is a large mastaba of the **usual shape**; … 'A sloping passage turns horizontal at the bottom, passes three slides for portcullises, and lastly opens into a **chamber running east and west**, with a ridge-roof. From the west end opens another chamber with a barrel-roof. And from the east end of the south side is a short horizontal passage, with four recesses and a small chamber. **The arrangement is closely like that of a pyramid**, and every part is equalled in that of Unas at Saqqara, though rather different arranged' (Petrie, History of Egypt, I, p.94). The mastaba had a funerary temple on its east side" (Baikie, 2018). "The entrance to the subterranean system of chambers is located on the shorter, northern side". https://en.wikipedia.org/wiki/Mastabat_al-Fir%27aun.

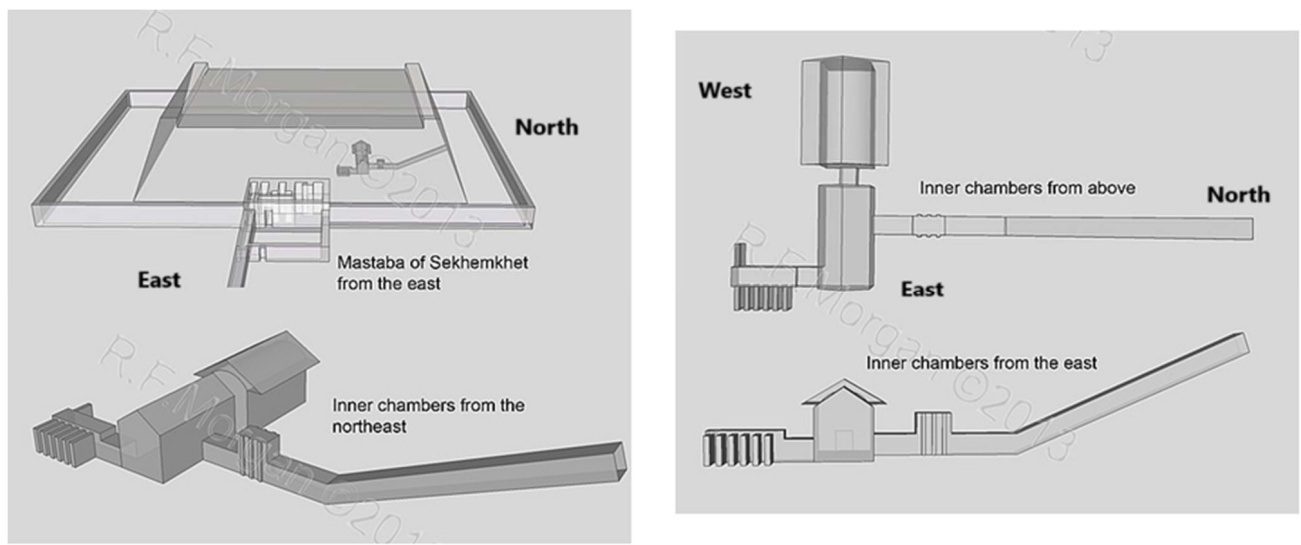

"The mastaba and close-up views of the inner chambers of Shepseskaf", Date 17 July 2013, 00:13:36. Author R.F.Morgan. https://commons.wikimedia.org/wiki/File:Shepseskaf_Iso.jpg

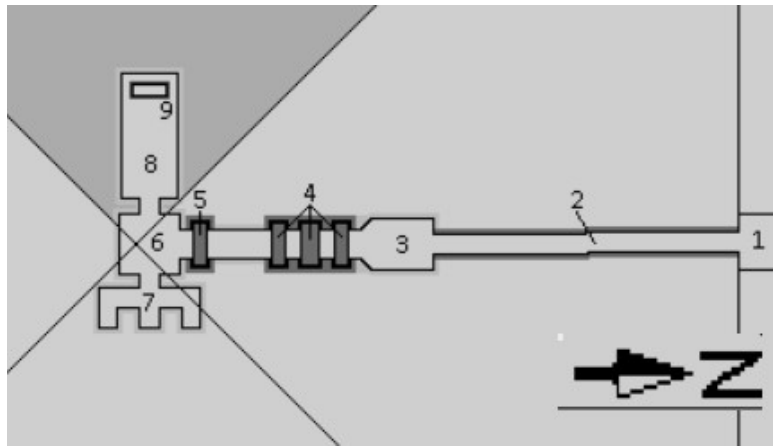

For comparison, the Djedkare Isesi Pyramid substructure.

Is Mastabat Fara'un a mastaba? No. It is a pyramid, because the substructure is that of a Pyramid.

In the following schematic view, a typical mastaba. It is composed of a subterranean structure, characterized by a **vertical pit, NOT by a corridor**.

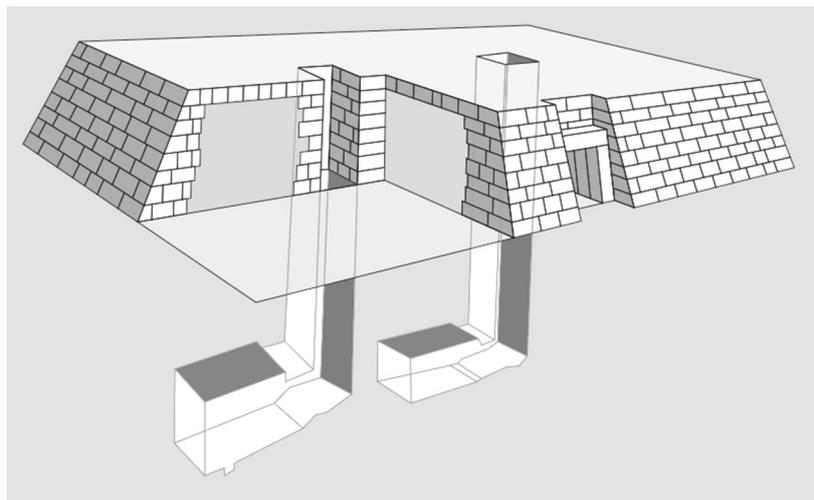

Image Courtesy Oesermaatra0069 , Master Uegly, for
https://commons.wikimedia.org/wiki/File:Mastaba_schematics.svg.

"Mastabas were … oriented north–south, which the Egyptians believed was essential for access to the afterlife" https://en.wikipedia.org/wiki/Mastaba This is true for Giza necropolis, but, at Abydos, we find earlier mastabas oriented following the Nile river.

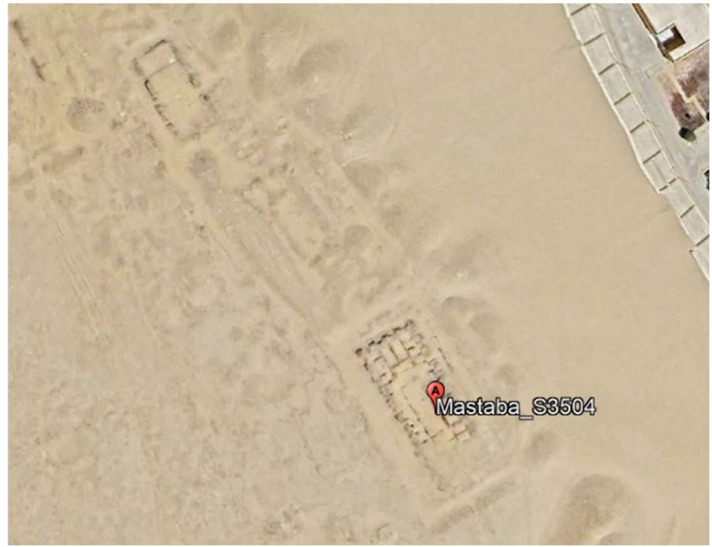

Mastaba S3504 Courtesy Google Earth

"Mastaba S3504 (Saqqara Tomb No. 3504) is a large mastaba tomb located in the Saqqara necropolis in Lower Egypt. It was built during the reign of the ancient Egyptian Pharaoh Djet, in the First Dynasty (Early Dynastic Period), shortly after 3000 BC. It is one of the largest mastabas from this dynasty. … The actual mastaba superstructure contained 43 chambers. Below this was the burial chamber, which was surrounded by additional store rooms. The burial chamber itself was originally clad in gilt wood." https://en.wikipedia.org/wiki/Mastaba_S3504

From https://egyptsites.wordpress.com/2009/02/19/mastaba-of-shepseskaf/ : "The tomb is constructed of enormous blocks of limestone and was originally sheathed in a finer white Tura limestone casing, with a bottom course of pink granite. Remains of restoration texts of Prince Khaemwaset have been found on some of the casing blocks. The mastaba appears to have been built in two steps and may have been deliberately conceived to take the shape of a Buto-type shrine, a Lower Egyptian form of tomb which was a vaulted shape with straight ends and which Karl Lepsius noted as looking like a giant sarcophagus" (egyptsites). Let me stress that the "Buto-type shrine" is an idea of Mark Lehner, that we find in his Compete Pyramids. No archaic temple in Buto has survived to this day. For a detailed discussion of Buto shrine see Abd-Alghafour et al. 2023.

"The tomb is entered by a sloping passage on its northern side, about one and a half metres above ground level and very similar to a pyramid entrance. This descends about 20m into a corridor originally blocked by three portcullis slabs and leads to the subterranean antechamber, burial chamber and store-rooms. The antechamber and burial chamber both have ceilings constructed as a false vault, like those in Menkaure's pyramid and both of the chambers were built with pink granite. The burial chamber contained fragments of Shepseskaf's dark basalt sarcophagus, but little else. From the antechamber a narrow passage runs to the south and leads to six niches or store-rooms" (egyptsites).

"The mastaba was enclosed within two mudbrick walls, the first containing Shepseskaf's mortuary temple on the eastern side. The small temple seems to have been constructed in two phases, the earlier parts in stone with later mudbrick additions. The older parts of the mortuary temple included a paved courtyard with an altar, a T-shaped offering hall with a false door and several chambers which were probably magazines. The later mudbrick parts had a large courtyard built to the east with niches decorating the inner walls" (egyptsites).

"Shepseskaf's causeway, constructed from white-painted mudbrick, adjoined the mortuary temple at the south-eastern corner of the courtyard wall. When built, the long causeway resembled a vaulted passage which must have led down to the King's valley temple but this has not yet been discovered" (egyptsites).

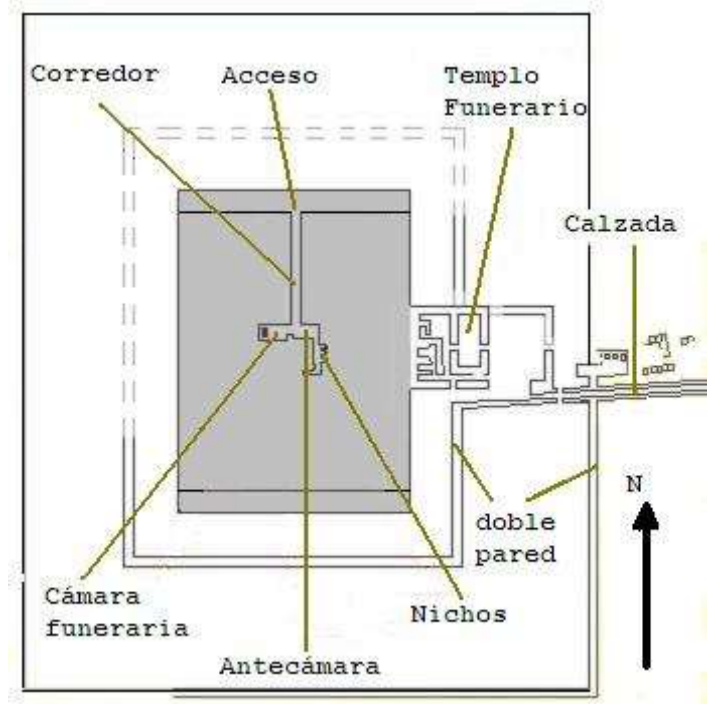

Groundplan of the tomb complex. Courtesy Gusgus (Plano del complejo funerario de Shepseskaf). CC BY 4.

https://en.wikipedia.org/wiki/Mastabat_al-Fir%27aun#/media/File:Mastabashepseskaf.jpg

**Hassan's Shepseskaf**

Hassan, 1943. "Now the pyramid was in itself perhaps the acme of materialism in one sense, for it may be regarded as a stupendous effort on the part of a nation to preserve the mortal body of one man by encasing it in the heart of a mountain of stone, even as a fossil is preserved in its matrix of surrounding rock, but regarded from another angle, the pyramid was not merely the geometrical form of a mass of stone erected over a royal grave, it was a sacred emblem, the symbol of the benben, the sun-stone of Heliopolis. To be buried within a pyramid was to be, as it were, merged into the very core of the holy symbol. To adopt a pyramid-tomb was to recognise the preeminence of the Solar cult, and to place one's hopes of a future life under the protection of the Sun-god Ra'. This idea had apparently been faithfully followed by his predecessors, why then, did Shepses-ka-f break away from the traditions of his family? It is possible that he was influenced in his religious ideas by the sheer force of public opinion. Did the overwhelming weight of the material cult, professed by the great mass of the populace, lead him to abandon the vaguer doctrine of the Solar-cult, and, turning his back on the pyramid-dominated necropolis of Giza, and all the ideas it stood for, erect his sarcophagus-shaped " House of Eternity " for his Ka, at Sakkara? Another possibility is that Shepses-ka-f adopted the material cult in order to gain a greater popularity with his subjects. … As we have already seen, the power of the Heliopolitan priesthood was steadily increasing throughout the whole time of the Fourth Dynasty, while on the contrary, the royal power, which had reached its zenith under Khwfw, was gradually, but none the less steadily, declining. This may be proved by the funerary monuments of the kings succeeding Khwfw, each of which in turn falls a little short of the splendour of its predecessor. Perhaps, then, Shepses-ka-f foresaw the inevitable result of the increasing aggrandizement of the Ra' priesthood, and forestalling the move of Akhenaton, some 1,370 years later, made a bold attempt to shatter the sacerdotal power by renouncing the accepted state religion, and making a clean break with all that it stood for. In the case of Shepses-ka-f, he seems to have completely abandoned the Solar-cult and adopted the material creed of the Ka. As proof of this we have: (1) His sarcophagus-shaped tomb (the sarcophagus being the very heart and core of the eternal dwelling of the Ka), and furthermore, **orientated this tomb to the east**, the direction in which the Ka was thought to enter and leave the burial-chamber. He abandoned the old necropolis, hallowed by those gigantic solar symbols, the pyramids, and chose a spot for his tomb far to the south, near Sakkara. He did not adopt the Ra' element in his name. Manetho hints at a short period of anarchy following the death of Men-kaw-Ra', but there is not a shred of evidence from the monuments to prove this statement; it may be, however, an echo of a fierce though bloodless battle waged between a far-seeing but not over-powerful monarch and a jealous and ever increasingly-powerful priesthood. That the feud (if it ever really existed) did not irrevocably blast the King's reputation is proved by the fact that the name of Shepses-ka-f is recognized as that of a legitimate monarch, in the King-lists, neither was his memory hated by either priests or people of the succeeding Solar Dynasty, and his name and monuments escaped destruction at the hands of the adherents of the Solar-cult. While in the later case of Akhenaton, the very reverse happened. " (Hassan, 1945).

Let us note that:

*The Shepseskaf tomb is oriented to the north.*

*Substructures are those of a pyramid.*

*The Snefru's Red Pyramid is also far from Heliopolis, it is a pyramid with the form of a benben.*

*Shepseskaf pyramid is near Dahshur pyramids.*

**The name of the pyramid**

"Shepseskaf's tomb is a great mastaba at South Saqqara. Called Qbḥ-Špss-k3.f ("Qebeh Shepseskaf") by the ancient Egyptians, this name is variously translated as "Shepseskaf is pure",[77] "Shepseskaf is purified", "Coolness of King Shepseskaf"[139] and "The cool place of Shepseskaf".[8]". See https://en.wikipedia.org/wiki/Shepseskaf and the references therein. [139] is Bogdanov, 2019.

According to Bogdanov, the name of the pyramid is given on the false door of Nikauhor tomb at Giza. http://giza.fas.harvard.edu/ancientpeople/2406/full/: "Owner (along with Ankhwedjes) of Service tomb 11. False door inscribed for Nikauhor, identified as [rx nswt sHD wabw jmj-r qbHw-SpsskAf] royal acquaintance, inspector of wab-priests, overseer of the pyramid of Shepseskaf; in situ in Service tomb 11." jmj-r overseer, qbH cool place.

Ahmed Fakhry, 1935, describes the tomb.

The Palermo stone tells us that the name of the pyramid was: 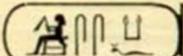

The sign used in the Nikauhor false door tells us that this pyramid had a nickname: "the sarcophagus".

For the Palermo Stone, see https://commons.wikimedia.org/wiki/File:Palermo_stone-verso_side.jpg (Neville, La pierre de Palermo, Libreairy Émile Bouillon, Paris 1903)

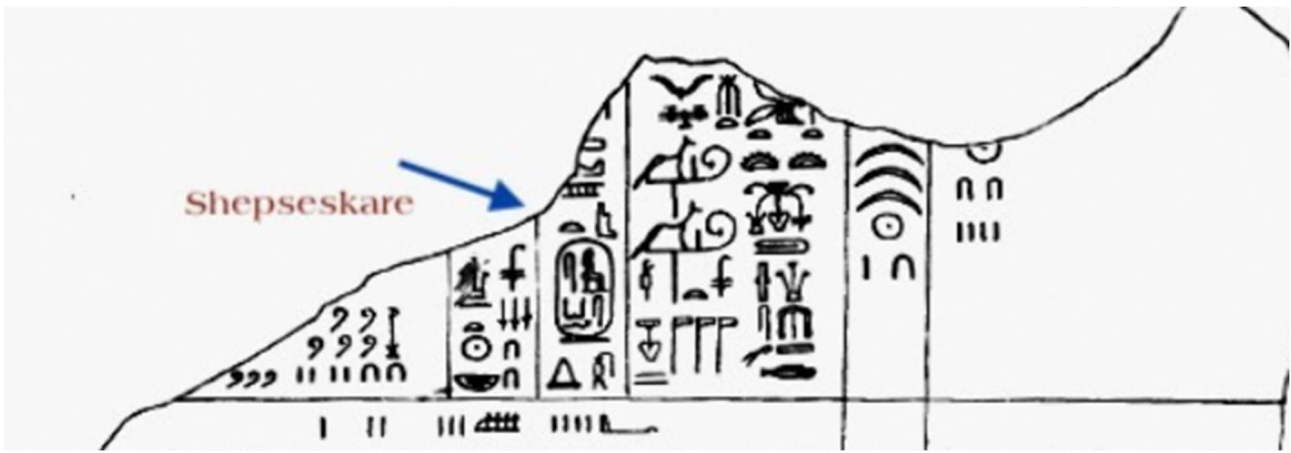

Owner (along with Ankhwedjes) of Service tomb 11. False door inscribed for Nikauhor, identified as [rx nswt sHD wabw jmj-r qbHw-SpsskAf] royal acquaintance, inspector of wab-priests, overseer of the pyramid of Shepseskaf; in situ in Service tomb 11. https://en.wiktionary.org/wiki/qbḥ, https://en.wiktionary.org/wiki/qbḥw#Egyptian , https://en.wiktionary.org/wiki/-w#Egyptian , https://en.wiktionary.org/wiki/jmj-r-mšʿ

| | G43<br>U+13171 | quail chick | **Field, district, region (w)** | w | 1. Unil. *w*<br>2. Either "quail chick" or equivalent coil (hieroglyph), Gardiner Z7, ℮<br>, used also for the *plural* word ending |
|---|---|---|---|---|---|
| | W16<br>U+133C2 | water jar with rack | Libation (qbḥ), (qbb) | | |

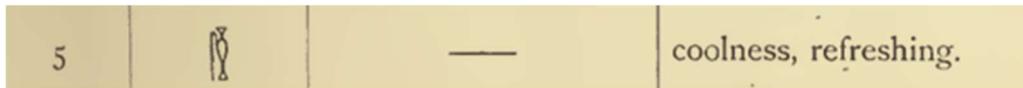 (Budge, 1981).

Nikauhor, overseer of the field of … .

https://web.archive.org/web/20201001060404/http://giza.fas.harvard.edu/ancientpeople/2406/full/

**The fence of Lebanon wood**

Bogdanov, 2019: "The Palermo fragment of the annals from the reign of Spss-kA=f. The partially destroyed column of the Old Kingdom annals in the part that deals with the events of the reign of Spss-kA=f contains the following text:

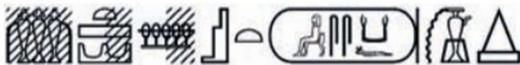 [...xn]tj-S Szp st qbH(w)-Spss-kA=f. Many researchers who have dealt with this text understood the term xntj-S as a title, but it is not appropriate in that context. Taking into account the lapidary nature of the annals, it just would not make any sense to mention the executors of the operation. If one understood xntj-S here as "Lebanese wood," the whole phrase would take on a completely satisfactory meaning: "[...] Lebanese wood enclosure of the place of the pyramid Coolness of king Spss-kA=f." Further arguments can be adduced for such an interpretation. It should be noted that the event occurred in the first year of the Spss-kA=f reign when the pyramid had not yet been built. Consequently, we are talking about the very beginning of construction, when the plateau where the pyramid was to be located was enclosed with a fence (Szp). The toponym qbH(w)-SpsskA=f with the determinative △ is known from a very small number of documents, and the plan for construction of a mastaba instead of a pyramid probably did not change immediately." Note 45 in Bogdanov, 2019: "BM EA 1234 (JaMes 1961, 11, pl. 11 (2)); BudGe 1896, 97 (41); cf. title jmj-rA qbHw-Spss- kA=f "overseer of the tomb Coolness of king Spss-kA=f " with the determinative "mastaba" in the tomb of n(j)-kAw-Hr (Giza; Service tomb 1; Fakhry 1935, 5, Fig. 2; www.gizapyramids.org: A7394_NS )", that is the tomb of Nikauhor at Giza.

Once more, the substructure of the Mastabat Fara'un is that of a pyramid, not of a mastaba. The project changed from pyramid to mastaba.

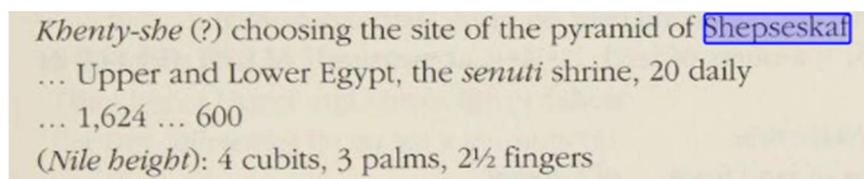

"Texts from the pyramid age", by Strudwick, Nigel

**Purified**

From Dickinson, 2014. "Shepseskaf, son and successor of Menkaure, was considered in later king lists the last king of the 4th dynasty. Shepseskaf returned to Saqqara for his RMC [Royal Mortuary Complex], which he placed south of the Saqqara plateau where Djoser and Sekhemkhet had their

RMCs built to extend the earlier royal necropolis south (figs. 2.17, 8.9.1-8.8.9.). Shepseskaf had a short reign; Manetho assigns him seven years and the Turin Canon four. Within the span of his short reign, Shepseskaf was responsible for the construction of a much smaller RMC and the completion of his father's RMC at Giza during the first year of his reign. Shepseskaf, which means 'His Soul (or Body) is Noble' and whose pyramid was named *'Shepseskaf is Purified' (Verner 2001: 254),* broke the royal funerary tradition by building a stone mastaba instead of a pyramid. This gave the complex its name, Mastaba Faraun; and it did not have a followers' cemetery. However, the overall layout of his RMC retained all the traditional features of a 4th dynasty RMC and the layout of his internal apartments actually set the norm for all subsequent OK [Old Kingdom] royal tombs (Goedicke 2000: 405). Vyse and Perring (1840), Mariette (1884: 361-5) and Lepsius (1943) give brief descriptions of the monument, but Jéquier (1928) carried out the first and only systematic excavation of the RMC. Maragioglio and Rinaldi's (1967) survey work provides additional information on the complex". About the location: "The desert surface is even and consists of a marly limestone (Klemm and Klemm 2010: 109) that is moderate to poor in quality. The nature of terrain required artificial foundations to be built and may explain in part the reduced scale of the superstructure (Jéquier 1928:3-4; Maragioglio and Rinaldi 1967: 136)." (Dickinson, 2014, and references therein).

"Shepseskaf leaves Giza for South Saqqara, ushering an age of capital-based RMCs (Abusir and South Saqqara). He is best known for his abandonment of the pyramid shape. Still, much in the design of his RMC continues previous norms and shows signs of skill and expertise with limited numbers of workers. While Shepseskaf's move is generally seen as ideological, moving away from the dominance of the sun cult, the move may in fact have been logistical, prompted by the retreat of the Nile east at Giza (Bunbury et al. 2009: 163). This left limited space available on the plateau to build another RMC. Yet Shepseskaf had a desire to be closer to the capital, which may then have already been undergoing a southern expansion (Malek 1997). It is also significant that mudbrick remained an important material for his RMC, but also those of his successors, until the end of the 4th dynasty. Khufu and Khafre's RMCs at Giza were the only exceptions." (Dickinson, 2014, and references therein).

See the text at https://discovery.ucl.ac.uk/id/eprint/1448710/7/DICKINSON%20PHD%20V1.pdf

https://web.archive.org/web/20250216150647/https://discovery.ucl.ac.uk/id/eprint/1448710/7/DICKINSON%20PHD%20V1.pdf

**Hawass' Shepseskaf**

From the "Mountains of the Pharaohs", 2006, by Zahi Hawass. "Shepseskaf ruled for only about four years and built a large mastaba, a low rectangular tomb, rather than a pyramid. The last ruler of the 4th Dynasty, a queen named Khentkawes (c. 2519-2513 B.C.), returned to Giza to build a massive tomb, but this was the last great monument built here. Afterward, the site was abandoned in favor of other pyramid fields to the south, although its royal cults continued to flourish for hundreds of years. Giza was the heart of Egypt for three generations." (Hawass, 2006). "The mastaba **is in the form of a giant sarcophagus**, with rectangular sides and an arched roof. Like the pyramids of his forebears, Shepseskaf s mastaba represents the primeval mound on which the creator god stood to bring the cosmos into being. However, his tomb does not carry the blatantly solar overtones of the true pyramids, and scholars have questioned his relationship to the cult of the sun god … It may be simply that he did not have the wealth needed to build a pyramid. He seems to have continued with the political systems of his father and in most ways continued the traditions of his 4th Dynasty predecessors. In addition to finishing his father's complex, he honored Menkaure's cult by exempting the priests of the temples from paying taxes. In other ways, Shepseskaf did break with ancient tradition. He married his daughter to a man, Shepsesptah, who was not of royal blood but lived at the palace of Menkaure. This was the first time in pharaonic history that we know of a royal princess marrying a commoner." (Hawass, 2006).

**The last King of the Fourth Dynasty, Queen Khentkawes**

Who was "Khentkawes, a mysterious female monarch who declared in her own colossal monument at Giza that she was the mother of either one or two kings of Upper and Lower Egypt? Her titles may imply that she ruled Egypt as king in her own right at the end of the Giza dynasty, so it is to this elusive queen that we next turn our attention. A massive monument, in the shape of a giant square sarcophagus on a high podium like that of Shepseskaf, still stands at the south of the Giza plateau between the causeways of Khafre and Menkaure. The base of the tomb, which was cut into the solid rock and cased with fine white limestone, measures forty-five meters (148 feet), and its original height was forty-five meters (148 feet). The size and square base of this tomb prompted its main excavator, Selim Hassan, to christen it the "fourth Giza pyramid." Under the tomb is a burial chamber lined with granite and containing a false door, along with seven small rooms designed to house furniture to be used in the afterlife. A small mortuary temple stands against the eastern face of this structure, and a causeway leads to a valley temple." (Hawass, 2006).

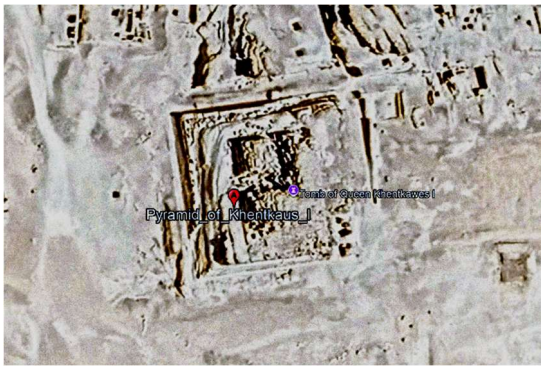 The 'fourth pyramid' at Giza, Courtesy Google Earth.

"The mortuary temple was cased with granite and decorated with religious scenes. On the south-west side of the tomb, Hassan found a boat pit cut into the rock; matching this to the northeast is a rectangular pool entered by eleven steps. This tomb is dedicated to a queen named Khentkawes, who was a daughter of the god and also claimed a title that can be read as either "king's mother and king of Upper and Lower Egypt" or "mother of two kings of Upper and Lower Egypt." Thus, she may have been a ruling queen, the last monarch of the 4th Dynasty. She was certainly a princess; although her parentage is technically unknown, it is most likely that she was the daughter of Menkaure and the sister and wife of Shepseskaf, giving her multiple claims to the throne in the absence of a strong male heir." (Hawass, 2006).

"As a woman, Khentkawes would not technically have had the right to hold the throne of Egypt, and, not surprisingly, her name does not appear in any of the later king lists. However, her clearly royal monument and her title suggest strongly that however ephemerally, she sat on the throne of the Two Lands. One image of the queen in her tomb, recently examined by Czech Egyptologist Miroslav Verner, shows her wearing a uraeus cobra on her brow, normally seen only on kings during this period. Within her valley temple, a small pyramid city thrived, demonstrating that her cult was active for a significant period of time and providing further evidence that she was accepted as a ruler of Egypt." (Hawass, 2006).

In fact, in the https://en.wikipedia.org/wiki/Abydos_King_List, the queen Khentkawes (or Khentkaus I) is not present.

As the last king of the IV Dynasty, we find sometime mentioned king Djedefptah (Thamphthis). "Winfried Seipel and Hermann Alexander Schlögl instead postulate that the historical figure behind Thamphthis could have been queen Khentkaus I.[6] This theory is supported by Khentkaus being depicted in her mortuary temple as a ruling pharaoh with nemes-headdress, king's beard and uraeus-diadem on her forehead. But this theory is problematic since Khentkaus' name never appears inside a serekh or royal cartouche.[7]" (https://en.wikipedia.org/wiki/Thamphthis, mentioning [6] Seipel, 1980, and [7] Schlögl, 2006).

"Some scholars believe that Khentkawes was the mother of Userkaf, first ruler of the 5th Dynasty, and thus provides a link between the end of the 4th Dynasty and the beginning of the 5th. It has also been theorized that Khentkawes married a priest of Heliopolis and that her children and the heirs to the throne were only half royal." (Hawass, 2006).

**From the IV to the V Dynasty**

"**Texts in several tombs at Giza indicate that there was no dramatic political break between the 4th and 5th Dynasties**. One of Khafre's sons, Sekhemkare, recorded in his tomb that kings Khafre, Menkaure, Shepseskaf, Userkaf, and Sahure all paid him honor. … **We must remember that the whole concept of dynasties is a later construct, that the Egyptians themselves saw their kings in an unbroken line**. However, there must have been some change of family, and the anomalous rule of Queen Khentkawes I provides evidence of difficulties with the succession" (Hawass, 2006).

Let us see the orientation of entrance and burial chambers of some of the pyramids of the Third Dynasty. Note that the Djoser pyramid is quite intricate inside.

**Djoser pyramid**

It is often told that the first pyramid was the Djoser pyramid. The "Djoser's pyramid is made up of six layers and was originally built as a type of rectangular tomb known today as a mastaba (an Arabic word meaning "bench") before being expanded into a step pyramid." (Jarus, 2023).

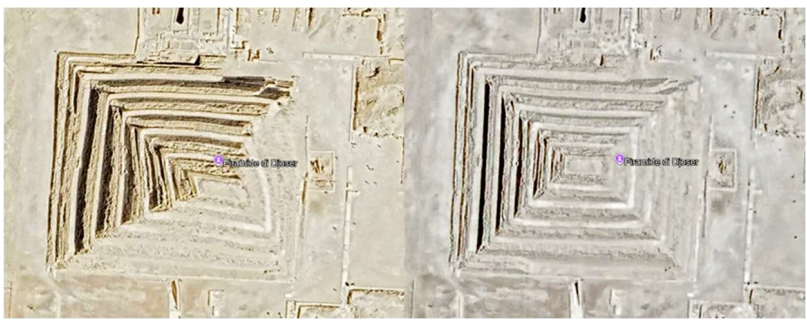

The default image Courtesy Google Earth on the left is giving an 'artistic' view of the monument. It seems the pyramid is ruining downside.

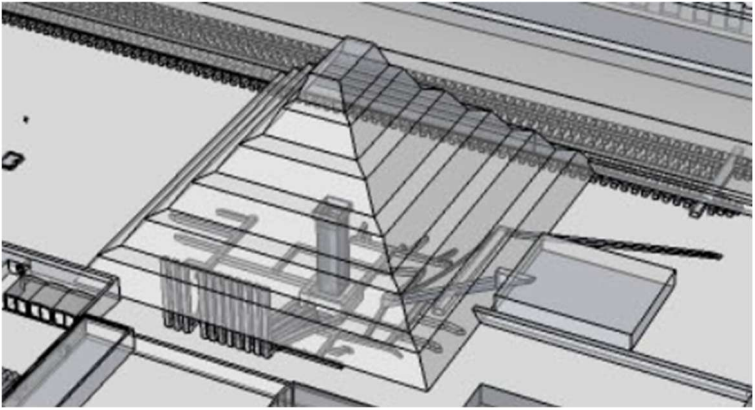

Inside Djoser pyramid. Courtesy R.F. Morgan, Isometric, plan and elevation images Djoser's Pyramid Complex taken from a 3d model. Available at:

https://en.wikipedia.org/wiki/Pyramid_of_Djoser#/media/File:001_Djoser.jpg

"Under the step pyramid is a labyrinth of tunneled chambers and galleries that total nearly 6 km in length and connect to a central shaft 7 m square and 28 m deep.[43] These spaces provide room for the king's burial, the burial of family members, and the storage of goods and offerings. The entrance to the 28 m shaft was built on the north side of the pyramid, a trend that would remain throughout the Old Kingdom. The sides of the underground passages are limestone inlaid with blue faience. … Together these chambers constitute the funerary apartment that mimicked the palace and would serve as the living place of the royal ka." https://en.wikipedia.org/wiki/Pyramid_of_Djoser , [43] is Verner, 1998.

**Sekhemkhet (Djoserty) pyramid, that is the "buried pyramid"**

"Sekhemkhet's pyramid is sometimes referred to as the "Buried Pyramid" and was first excavated in 1952 by Egyptian archaeologist Zakaria Goneim. A sealed sarcophagus was discovered beneath the pyramid, but when opened was found to be empty." "The burial chamber has a base measurement of 29 ft x 17 ft and a height of 15 ft. It was also left unfinished, but surprisingly a nearly completely arranged burial was found. The sarcophagus **in the midst of the chamber** is made of polished alabaster and shows an unusual feature: its opening lies on the front side and is sealed by a sliding door, which was still plastered with mortar when the sarcophagus was found. The sarcophagus was empty, however and it remains unclear whether the site was ransacked after burial or whether King Sekhemkhet was buried elsewhere." https://en.wikipedia.org/wiki/Sekhemkhet

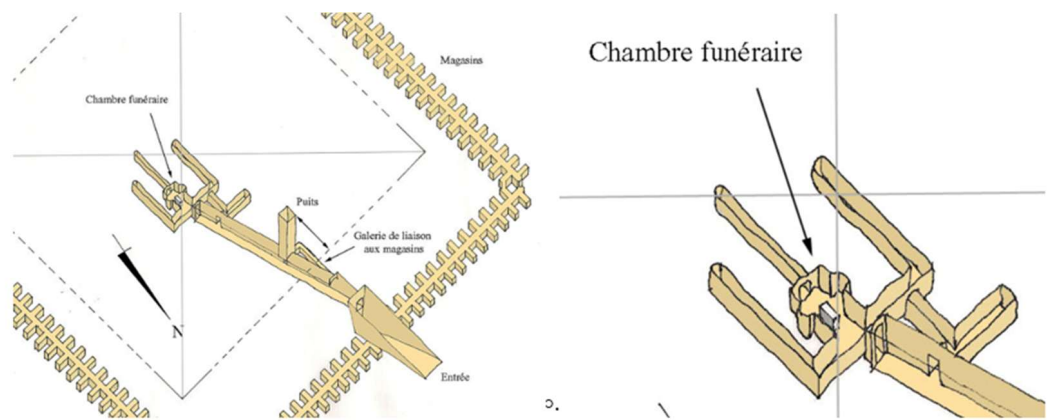

« Vue axonométrique des infrastructures de la pyramide de Sekhemkhet à Saqqarah, 7 July 2007, Courtesy Franck Monnier » .

https://commons.wikimedia.org/wiki/File:Infrasctructure-sekhemkhet.jpg

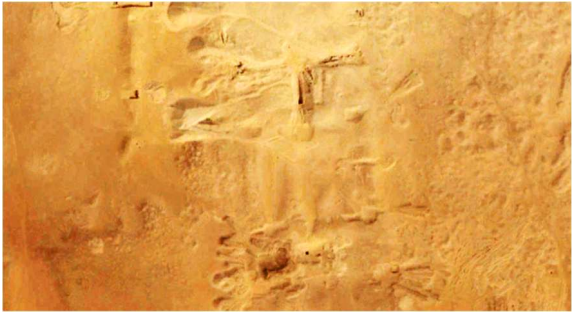

The site of the buried pyramid, Courtesy Google Earth.

"The Buried Pyramid (also called the Pyramid of Sekhemkhet) is an unfinished step pyramid constructed c. 2645 BC for Sekhemkhet. He was the second king of the Third Dynasty of Ancient Egypt, which reigned over Egypt c. 2686–2613 BC and is usually placed at the beginning of the Old Kingdom of Egypt. Many historians believe that the third dynasty played an important role in the transition from Early Dynastic Period of Egypt to the Age of the Pyramids." https://en.wikipedia.org/wiki/Buried_Pyramid "When the blocked wall was breached, on May 31, 1954, an unfinished and undecorated burial chamber was discovered. Inside it, lay an alabaster sarcophagus cut from a single block with a vertical lid which seemed to still be sealed. On June 26, 1954, after great difficulties to unblock and raise the lid, the sarcophagus was opened and was found to be empty." Wikipedia mentions Lehner,

From the plans given above, we can find a north entrance. The sarcophagus seems to have been north oriented. "This box and lid was cut from a single piece of translucent alabaster stone, it has only 1 opening. And is unique compared to all other sarcophagus examples found in Egypt because it has a sliding door located on the side, rather than a lid on the top."

https://www.facebook.com/MegalithResearch/posts/the-mysterious-sarcophagus-from-sekhemkhets-unfinished-pyramidthis-box-and-lid-w/734200916701681/

### Khaba pyramid

"The Layer Pyramid (known locally in Arabic as el haram el midawwar, Arabic: الهرم المدور, meaning 'rubble-hill pyramid') is a ruined step pyramid dating to the 3rd Dynasty of Egypt (2686 BC to 2613 BC) and located in the necropolis of Zawyet El Aryan. Its ownership is uncertain and may be attributable to king Khaba. The pyramid architecture, however, is very similar to that of the Buried Pyramid of king Sekhemkhet and for this reason is firmly datable to the 3rd Dynasty." https://en.wikipedia.org/wiki/Layer_Pyramid

See the plan at https://en.wikipedia.org/wiki/Layer_Pyramid#/media/File:Khaba_1.jpg

### Huni pyramid

"Huni's burial site remains unknown. Since the Meidum pyramid can be excluded, egyptologists and archaeologists propose several alternative burial sites. As already pointed out, Rainer Stadelmann and Miroslav Verner propose the Layer-pyramid at Zawyet el-Aryan as Huni's tomb, because they identify Huni with Khaba, who is in turn well connected with the Layer-pyramid, since several stone

bowls with his Horus name were found in the surrounding necropolis."
https://en.wikipedia.org/wiki/Huni#Edfu_South_Pyramid

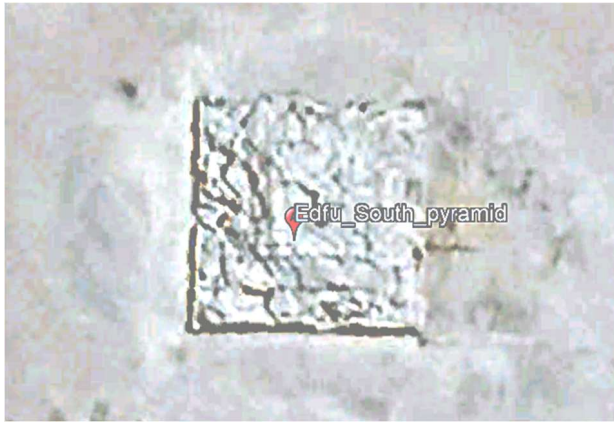 Edfu south pyramid, Courtesy Google Earth.

Djoser was the first king of the Third Dynasty, the last King was Huni. "The purpose of the [Edfu south] pyramid is unknown. Dreyer and Kaiser thought that it and the other pyramids named above were part of a single building project of Pharaoh Huni, the last ruler of the Third Dynasty. Andrzej Ćwiek mostly agrees, but suggests that Huni's successor, Sneferu (c.2670–2630 BC) the founder of the Fourth Dynasty, was the builder". https://en.wikipedia.org/wiki/Edfu_South_pyramid

**Before the Djoser pyramid**

The last king of the Second Dynasty was Khasekhemwi. His tomb is Tomb V at Umm el Qa'ab in Abydos. "Its trapezoid shape measuring 68.97 metres long by 10.04 metres on its shortest side and 17.06 metres on its longest side, makes it stand apart from the other royal tombs at Umm el-Qa'ab. It consists of a centrally located burial room constructed of dressed limestone blocks. It is surrounded by several smaller, inter-connecting chambers with mudbrick walls, that were probably used for storage. Unlike the other royal tombs at Umm el-Qa'ab, Khasekhemwi's tomb has two entrances, one in the north, the other in the south." https://www.ancient-egypt.org/

"On the surface, tumuli probably marked the spots of the single burial chambers, of the royal as well as the subsidiary, although no traces of such tumuli remained. The tumuli – of probably varying shape and dimensions – were made of sand, but might have had a mud brick casing, the tomb of Khasekhemwy even a layer of limestone blocks around the tumulus (fig. 7)." (Engel, 2008).

"Taking the varying geological situation at the different sites into consideration, the graves follow the same development for their subterranean construction: first, there are large chambers arranged in a row, followed by a central burial chamber with attached magazines. With the introduction of a staircase, it becomes an element at the different sites: the staircase first ends at the shorter side of the main burial chamber, in the next stage at the broad side, and finally, at the end of Dynasty 1, again at the smaller side. The development can be continued into Dynasty 2 since the tomb of Hetepsekhemwy expands the final ground plan employed for the tomb of Qa'a on a much larger scale (without the subsidiary chambers). While for Ninetjer and the unknown owner of Tomb C at Saqqara indeed a differing plan was used, Peribsen's tomb at Abydos copies those of the early Dynasty 1.

Khasekhemwy also takes this layout as a starting point, but had his tomb changed into something that resembled that of [Hetepsekhemwy](Hetepsekhemwy) at Saqqara" (Engel, 2008).

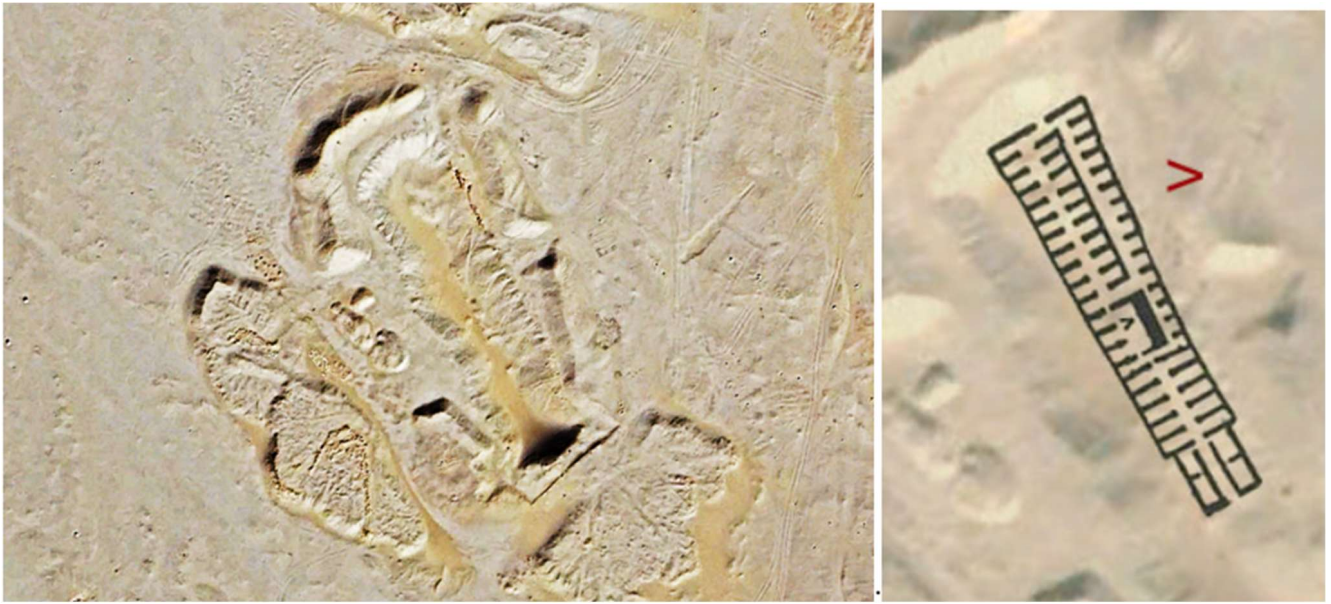

Coordinate 26°10'21.62"N, 31°54'29.12"E. On the left, image Courtesy Google Earth. On the right, Courtesy NASA

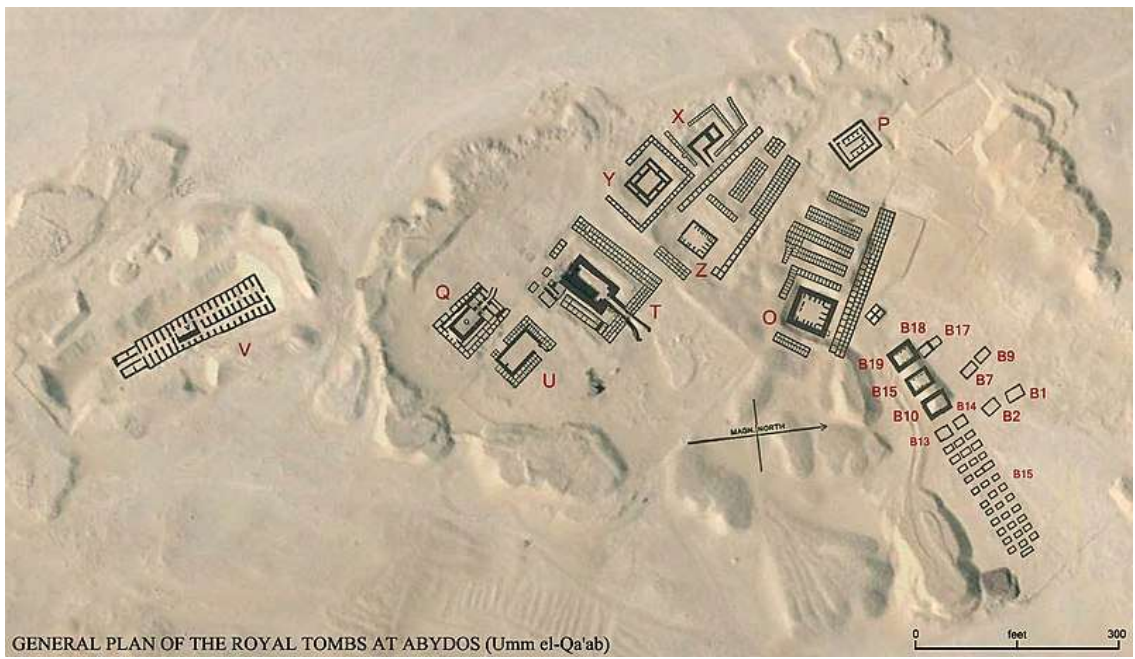

https://commons.wikimedia.org/wiki/File:AbydosSatMap.jpg

"Map overlay of Peqer necropolis at Umm el-Qa'ab near Abydos in Egypt. The tombs of pharaohs are marked with letters. Tomb of Iry-Hor (B1-B2) , Tomb of King Ka (B7-B9).

1st Dynasty

Tomb of Narmer (B17-B18)      Tomb of Aha (B19-B15-B10)      Tomb of Djer (O)

| Tomb of Djet (Z) | Tomb of Merneith (Y) | Tomb of Den (T) |
| Tomb of Anedjib (X) | Tomb of Semerhket (U) | Tomb of Qaa (Q) |

2nd Dynasty

| Tomb of Peribsen (P) | Tomb of Khasekhemwy (V) |

**The body**

We have information about the orientation of the body thanks to the article written by Raven, 2005.

The first phenomenon to be studied – says Raven – is the positions of the bodies in Egyptian burials. "Two main periods should be distinguished, the first during which the body of the deceased was laid on its side (almost always the left side, first in crouching position and later extended), the second when it lay on its back. The shift between the two traditions was gradual and took place between the Twelfth and Eighteenth Dynasties. In the earlier Predynastic cultures, the head was usually directed to the south, so that the face looked west." (Raven, 2005).

However, from the Later Predynastic Period onwards the opposite positioning with the head towards the north became increasingly common, and this remained the usual position throughout the Old Kingdom, First Intermediate Period, and Middle Kingdom. From the time when the body was laid on its back, however, the favorite position became that with the head towards the west, … the axis of the burials lay at right angle to that of the earlier ones. … Every cemetery had its exceptions from the general rule. These can often be explained by the local situation … It is also clear that the rules were obeyed more strictly in the richest tombs involving real burial changes …" (Raven, 2005).

It seems that there existed two different sets of symbolic connotations: one regarded the face, the other the body. "The direction towards which the deceased faced was considered to be of primary importance". In the predynastic burials, the face looked west, whereas, in the dynastic burials, they looked east. "The change of body direction at the end of the Middle Kingdom was a result of the introduction of the anthropoid coffin and the resulting supine position of the corpse. The prevailing east-west axis would enable the deceased to face east – if he raised his head – just as those buried in the rectangular coffins of the preceding period. A solar connection seems obvious, and a prehistoric orientation to the realm of the setting sun and the dead would then have given way to a pharaonic fixation on the rising sun with the promise of resurrection" (Raven, 2005).

Besides the orientation of the face, another orientation symbolism existed, that of the body as a whole. "Egyptian burials were positioned on a north-south axis for two millennia (that is, before the system changed with the introduction of the supine deposition and the mummiform coffin, around 1800-1500 BC). During the Prehistoric Period, the favorite position was that with the head to the south. During the whole period from the first pharaohs to the end of the Middle Kingdom, the favorite orientation of the body was that with the head towards the north. That the body positioning was associated with chthonic (Osirian) notions is suggested by the iconographic system of the ancient Egyptian coffin" (Raven, 2005).

**The divine kings and their tombs**

In van den Dungen, https://www.sofiatopia.org/maat/wenis.htm , we find reported from Lamy, 1981 "In the Neolithic period the dead were deposited in oval graves in fetal position, with the head at the south. In Lower Egypt the deceased was placed on his right side, his face turned towards the east, while in Upper Egypt, as all along the upper Nile, the dead person was placed on his left side, looking west. Often the body was wrapped in a cloth or an animal skin, the head resting on a cushion."

"On the threshold of the First Dynasty (ca. 3000 - 2900 BCE), the graves of the rulers and the élite consisted of neat mudbrick boxes, sunk in the desert and divided, like a house or an imitation palace, into several rooms. The tombs of the first kings followed this pattern, but with increased complexity. Situated far out in the desert near the cliffs at Abydos, they were marked by a pair of large stelæ and covered by a mound" (van den Dungen).

"With the arrival of the institution of kingship & sacred language, the royal ritual and its funerary cult came into existence. The kings of the First Dynasty were buried at Abydos (the cult place of Osiris), an indication of the Upper Egyptian origin of the Egyptian state … The institution of kingship was already strong & powerful" (van den Dungen).

"The superstructures of the first royal tombs at Abydos were simple mounds of sand held in place by a mudbrick revetment. Scholars conjecture the burial mound recalls the primeval mound which emerged from the waters at the time of creation. The mound is Solar, and refers to the first ray of Re shining on the first day after the waters receded. In the tombs of kings Den and Adjib (First Dynasty, ca. 3000 - 2800 BCE), the entrance stairway approaches the burial-chamber from the East and the rising Sun. The symbolism speaks for itself. Like the rising Sun, the king rose to the sky." (van den Dungen).

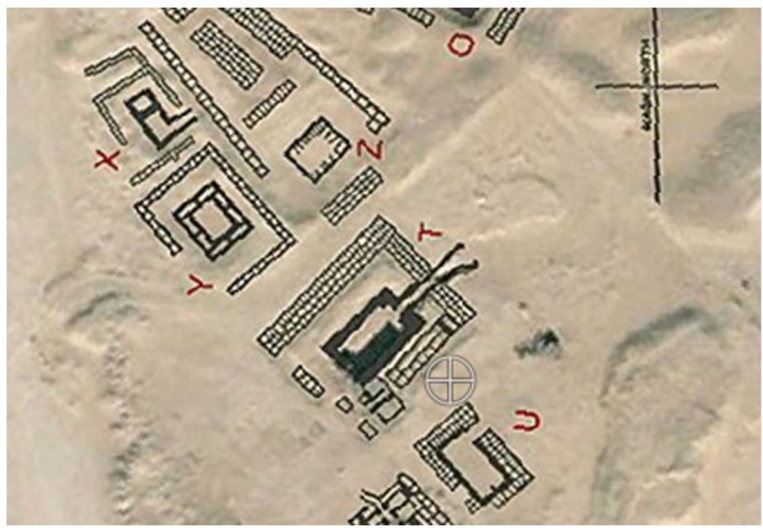

Tomb of Den (T), Tomb of Anedjib (X)

"Adjib's burial site was excavated at the Umm el-Qa'ab necropolis in Abydos and is known as "Tomb X". It measures 16.4 x 9.0 metres and is the smallest of all royal tombs in this area. Adjib's burial chamber (7 x 4.5 metres), consists of two rooms and is accessed by a stairway from the east. … The main chamber is surrounded by 64 subsidiary tombs which are interpreted as ancillary burials. Some of these chambers contained large numbers of ivory carvings." https://en.wikipedia.org/wiki/Anedjib

"In the tomb of King Qaa, who closes this dynasty, a change to a northerly orientation is effectuated (and maintained thereafter). The entrance corridor is a large ramp pointing northwards toward the circumpolar stars ("ixmw-sk" or "the ones that know not destruction"). **Funerary ideology became stellar**. The pyramids reflect a stellar ideology made possible by the local horizon delimiting the cycle of the Sun. They are made to assist the divine king on his celestial voyage to the stars" (van den Dungen).

"In these theological considerations, the change from mastaba to step pyramid, from primordial mound (of the Sun) to celestial ladder (to the stars), reflects the increased importance of the celestial, stellar terminus of the divine king. … The royals were divine beings, and so bound to the sky, whereas commoners hid beneath the Earth, in the dark kingdom of Osiris" (van den Dungen).

"Although the tombs left by the kings of the Early Dynastic Period are monumental in size, they do not approach the scale suddenly reached in the IIIth Dynasty (ca. 2670 - 2600 BCE), in particular under King Netjerikhet or "Djoser" (ca. 2654 - 2635 BCE) and his grand architect Imhotep." (van den Dungen).

Then the author passes to the Snefru and Khufu pyramids. "What is typical for these "Stellar" pyramids of Sneferu (Bent Pyramid as well as North Pyramid) and Khufu is the elevated position of the King's Chamber. In both, the funerary symbolism is clearly celestial. The expanse of the sky was the celestial Nile, with banks on the West and on the East." (van den Dungen).

**The Giza pyramids and the sun**

In 2016, I proposed https://arxiv.org/abs/1604.05963 , 'Khufu, Khafre and Menkaure Pyramids and the Sun'. In this paper, it has been "discussed the orientation of the Egyptian pyramids at Giza with respect to sunrises and sunsets, using this http URL software. We can see that Khufu and Khafre pyramids had been positioned in a manner that, from each pyramid, it was always possible to observe the points of the horizon where the sun was rising and setting on each day of the year".

The sarcophagus has its axis north-south. The body of the deceased king was lying on its left side, extended, head north, face towards the east. The king was looking at the sunrise, round the year.

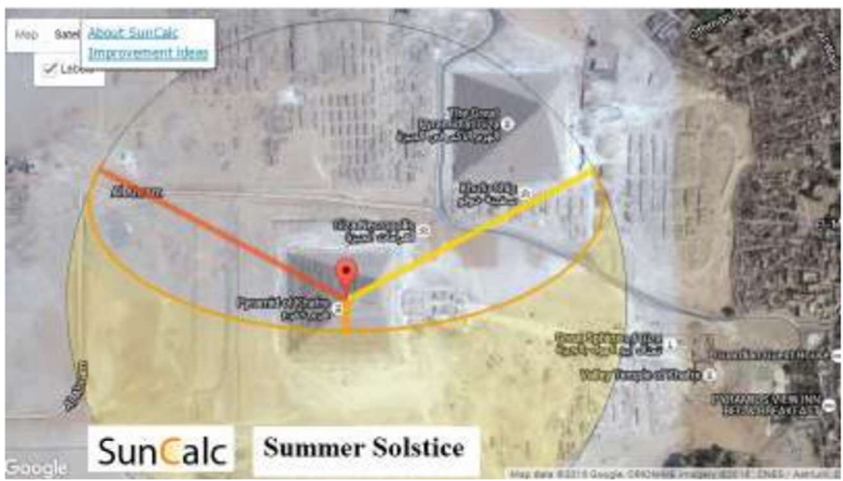

Khafre's architects have considered that their king would like to observe the sunrise round the year. Using SunCalc we can find that Khafre can see the point of the horizon where the sun is rising on the summer solstice, and then he can see the same on all the days of the year. No problems for sunset, because horizon was free. See also the Summer Solstice, https://www.suncalc.org/#/29.976,31.1307,15/2025.06.21/12:56/1/1

The same is true for Menkaure, as we can see for the Summer Solstice, https://www.suncalc.org/#/29.9724,31.1283,15/2025.06.21/12:56/1/1

**Pepi II (VI Dynasty) and Shepseskaf (IV Dynasty)**

"The pyramid of Pepi II was the tomb of King Pepi II, located in southern Saqqara, to the northwest of the Mastabat al-Fir'aun. It was the final full pyramid complex to be built in Ancient Egypt. … The substructure is similar to that of the Pyramid of Djedkare-Isesi, which was the model for all subsequent pyramids. A passageway descending from a point on the north face of the pyramid which was originally protected by the north chapel runs a little over twenty metres. It is blocked by four granite blocks at the entrance and leads to an entrance hall with a ceiling painted with five rows of white stars on a black background, oriented to the west. Then there is a horizontal corridor, itself blocked off by three granite blocks. The walls of this corridor are decorated with Pyramid Texts. It is followed by a funerary antechamber on an east–west axis and located under the very centre of the pyramid. The burial chambers were by a vault of eighteen massive stone blocks, arranged in chevrons. The ceiling of this vault was painted blue and covered with golden stars. On the eastern side of the antechamber, a doorway led to the serdab of the pyramid which has been completely destroyed. The burial chamber, whose walls are covered in pyramid texts, is 3.15 metres wide and nearly eight metres long (7.79 metres at the north end, 7.91 metres at the south end). The western wall of the chamber is painted with the facade of a palace. The sarcophagus is made out of greywacke; it is nearly three metres long, around 1.3 metres wide and 1.2 metres high. All four sides are engraved with hieroglyphs listing the complete royal titulary of Pepi II. The sarcophagus is a fine piece of work, but shows some traces of incompleteness with respect to the inscription, which also retains marks of preparatory guide lines and shows no signs of the gilding which was usual for a royal sarcophagus in this period. The lid of the sarcophagus is also made of greywacke and is more obviously unfinished; in places it was never smoothed and there are no traces of inscriptions. Some fragments of an alabaster chest for the canopic jars were found with the sarcophagus. The lid of this chest was also found, but is cut from a granite block - another sign of difficulty in completing the burial goods which were apparently completed in a hurry" https://en.wikipedia.org/wiki/Pyramid_of_Pepi_II

See here the plan of the substructure.

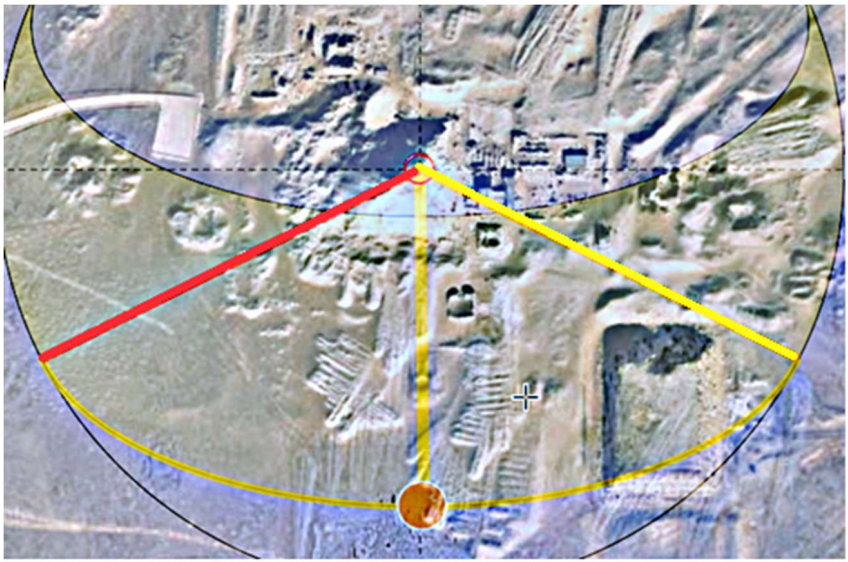

Sunrise and sunset on Winter Solstice. Pepi II can see the sunrise, round the year. Sunrise and sunset directions given by Suncalc.org

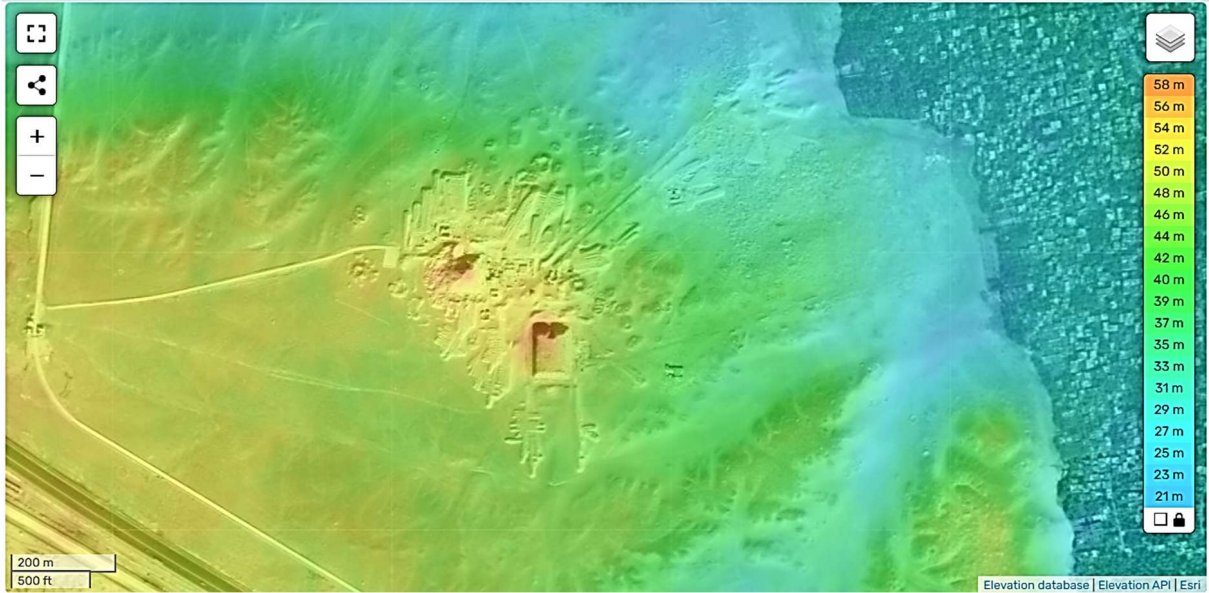

TessaDEM elevation map, Courtesy https://it-it.topographic-map.com/

### Neferirkare-Nyuserre (V Dynasty)

As you can see in the following image, we have the sunrise on the summer solstice linking two pyramids, that of Neferirkare Kakai and that of Nyuserre (on the east).

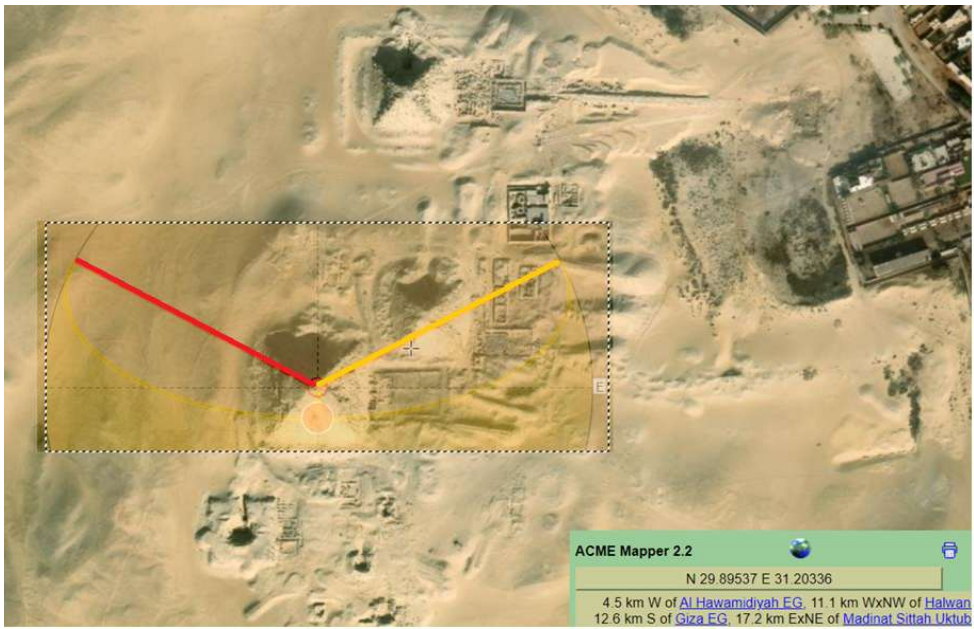

Abusir: sunrise and sunset directions of Summer Solstice given by Suncalc.org

"Nyuserre was the younger son of Neferirkare Kakai and queen Khentkaus II, and the brother of the short-lived king Neferefre. He may have succeeded his brother directly, as indicated by much later historical sources." https://en.wikipedia.org/wiki/Nyuserre_Ini

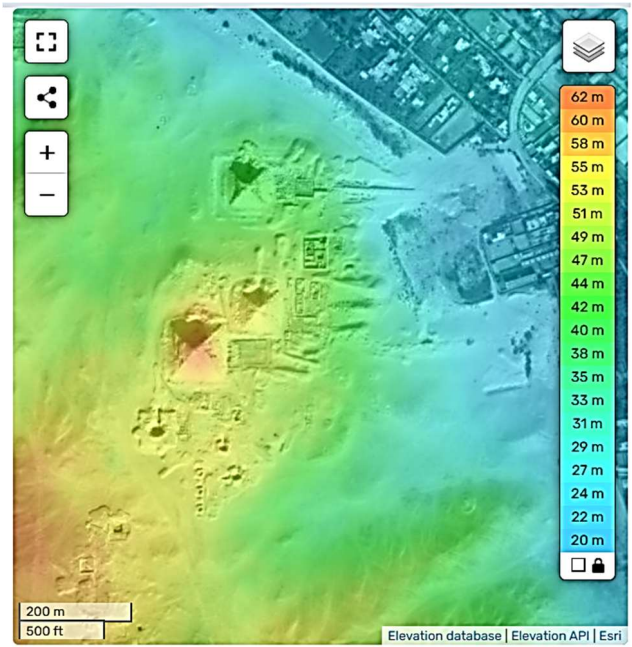 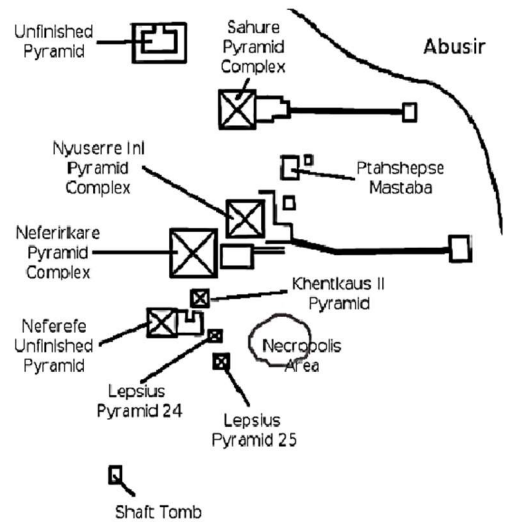

On the left, TessaDEM elevation map Courtesy https://it-it.topographic-map.com, on the right a map Courtesy to Commons by Maksim. https://commons.wikimedia.org/wiki/File:Abusir_map.png

Of course, the two pyramids are closely linked by the little space of this necropolis. Then, let us conclude by showing the case of Giza.

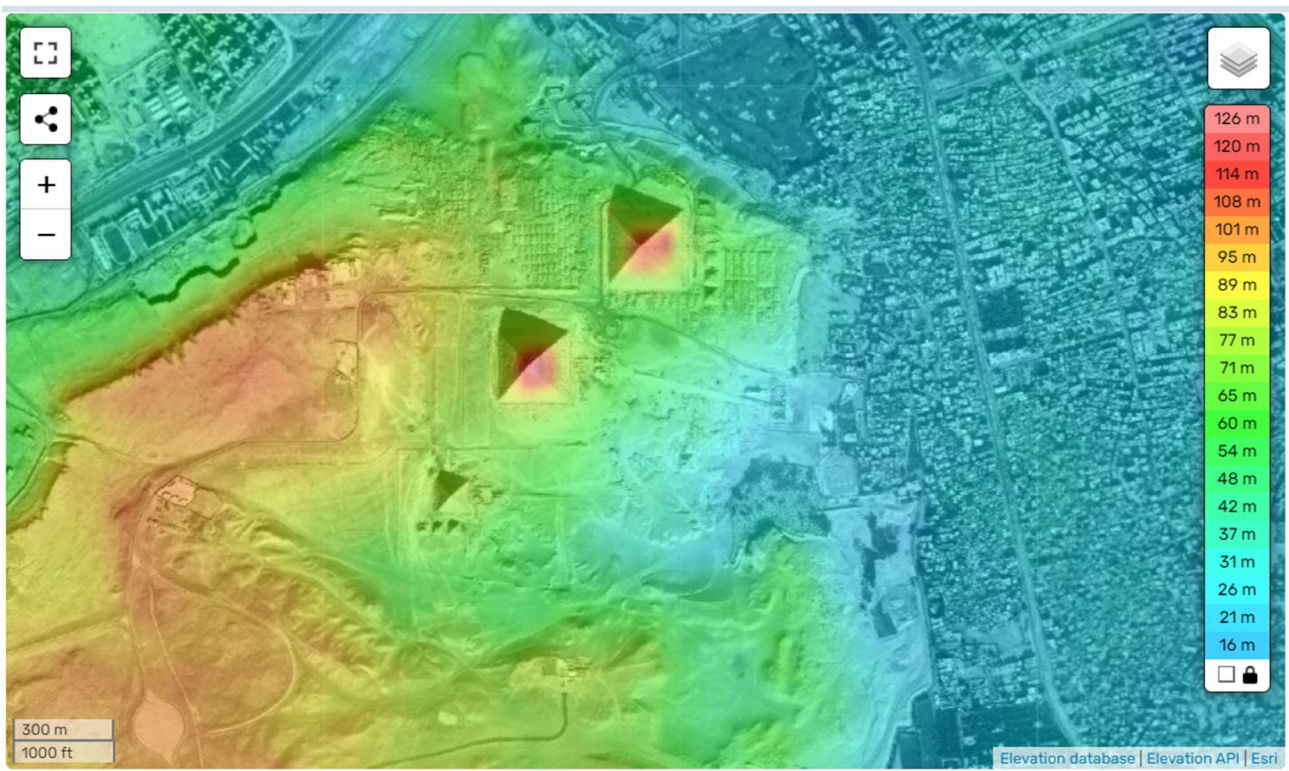

TessaDEM, Courtesy https://it-it.topographic-map.com/

Shepseskaf had no room to build his pyramid close to that of Khufu. He opted for a free space, close to the Snefru's Bent and Red pyramids. When possible, the pyramid builders had the opportunity to increase the perimeter of the burial site.